\documentclass[aps,twocolumn,nofootinbib]{revtex4}

\usepackage{graphicx}
\usepackage{dcolumn}
\usepackage{bm}

\usepackage{hyperref}
\hypersetup{
    colorlinks=true,
    linkcolor=blue,
    citecolor=blue,
    urlcolor=blue
}

\usepackage{amssymb,amsfonts}
\usepackage{epsfig}
\usepackage{epstopdf}
\usepackage[all]{xy}
\usepackage{amsthm}
\usepackage{dcolumn}
\usepackage{hyperref}
\usepackage{url}
\usepackage{dsfont}
\usepackage{slashed}
\usepackage{mathrsfs}

\usepackage{bbm}
\usepackage{tikz}
\usepackage{physics}
\usepackage{enumitem}
\usepackage{soul}
\usepackage{float}

\newcommand{\be}{\begin{equation}}
\newcommand{\ee}{\end{equation}}
\newcommand{\bea}{\begin{eqnarray}}
\newcommand{\nn}{\nonumber}
\newcommand{\eea}{\end{eqnarray}}

\newcommand*\interior[1]{\mathring{#1}}

\def\inbar{\,\vrule height1.5ex width.4pt depth0pt}
\def\IR{\relax{\rm I\kern-.18em R}}
\def\IC{\relax\hbox{$\inbar\kern-.3em{\rm C}$}}

\begin{document}

\title{Covariant Quantization of the Partially Massless Graviton Field in de Sitter Spacetime}

\author{Jean-Pierre Gazeau$^{1}$\footnote{gazeau@apc.in2p3.fr}}

\author{Hamed Pejhan$^{2}$\footnote{pejhan@math.bas.bg}}

\affiliation{$^1$Universit\'e Paris Cit\'{e}, CNRS, Astroparticule et Cosmologie, F-75013 Paris, France}

\affiliation{$^2$Institute of Mathematics and Informatics, Bulgarian Academy of Sciences, Acad. G. Bonchev Str. Bl. 8, 1113, Sofia, Bulgaria}

\date{\today}

\begin{abstract}
We present a covariant quantization scheme for the so-called ``partially massless" graviton field in de Sitter spacetime. Our approach is founded on the principles of the de Sitter group representation theory (in the sense given by Wigner), the Wightman-G\"{a}rding axioms for gauge-invariant fields (Gupta-Bleuler scheme), and the essential analyticity prerequisites in the complexified pseudo-Riemannian manifold. To implement the quantization process effectively, we adopt coordinate-independent (global) de Sitter plane waves. These plane waves, defined in the appropriate tube domains of complex de Sitter spacetime, serve as the de Sitter counterparts to the standard Minkowskian plane waves. By employing these analytical plane waves, we enable a spectral analysis of the corresponding two-point function that closely resembles the Fourier analysis typically employed in the flat Minkowskian case. Within this framework, we present the Wightman two-point function for the partially massless graviton field, which satisfies the essential criteria of locality, covariance, and normal analyticity. Furthermore, we provide insights into the underlying Hilbert space structure and the corresponding unsmeared field operator. A direct consequence of this quantization construction confirms the widely accepted notion of light-cone propagation for the de Sitter partially massless graviton field.
\end{abstract}

\maketitle

\setcounter{equation}{0} \section{Introduction}

\subsection{Motivation}
This paper is part of a series of papers/books (see Refs. \cite{Gazeau2022, Gazeau2023} and references therein) that attempts to develop a consistent formulation of elementary systems in the global structure of de Sitter (dS) and anti-dS (AdS) spacetimes, in the Wigner sense \cite{Wigner, Newton/Wigner}, as associated with unitary irreducible representations (UIRs) of the dS and AdS relativity groups, respectively;\footnote{To get the gist, let $P$ denote the physical systems whose global and local symmetries of their classical phase spaces are respectively determined by a Lie group $G$ and its Lie algebra $\mathfrak{g}$. [This is the case, for instance, for ``free" elementary systems living in dS and AdS spacetimes.] Then, the following statements hold:
\begin{itemize}
\item{The phase-space reading of $P$s can be realized by the orbits under the co-adjoint action of $G$ in the dual linear space to $\mathfrak{g}$ (traditionally, symbolized by $\mathfrak{g}^\ast$ in the literature). Such orbits, known as co-adjoint orbits, are symplectic manifolds. Moreover, each co-adjoint orbit carries a natural $G$-invariant (Liouville) measure and also, as a homogeneous space, is homeomorphic to an even-dimensional group coset $G/G_s$, where $G_s$, being a (closed) subgroup of $G$, stabilizes some orbit point. For more details, see Refs. \cite{Kirillov1,Kirillov2}.}
\item{Co-adjoint orbits, possessing very rich analytic structures, also underlie (projective) Hilbert spaces carrying UIRs of the respective symmetry group $G$. [Here, a comprehensive program of quantization of functions (or distributions) by considering all  references of covariant integral quantization (see, for instance, Refs. \cite{GazeauWiley,aagbook13,bergaz14,gazeauAP16,gazmur16}) can be taken into account.] In the sense that was initially put forward by Wigner \cite{Wigner, Newton/Wigner} in the context of Einstein-Poincar\'{e} relativity and then developed by others \cite{Wigner1952, Levy-Leblond, Voisin, Gursey1963, Fronsdal, Fronsdal'} to Galilean, dS, and AdS systems, the (projective) Hilbert spaces identify (in some restricted sense) quantum (``one-particle") states spaces in the respective quantum-mechanical reading of $P$s. The invariant parameters characterizing the (projective) UIRs then identify the basic quantum numbers characterizing the respective quantum states of $P$s. Remarkably, by construction, this quantization scheme guarantees a ``smooth" transition from classical to the quantum reading of the physical systems $P$.}
\end{itemize}
For a more detailed discussion, focusing on the dS and AdS cases, readers are referred to Refs. \cite{Gazeau2022,Gazeau2023} and references therein.\label{foot Wigner sense}} (A)dS relativity versus Einstein-Poincar\'{e} relativity. The motivation for this attempt, if we restrict our attention merely to the dS case which is of interest in the present study, is rooted in part in the key role that is played by the dS geometry in the \emph{inflationary cosmological scenarii}\footnote{According to the inflationary cosmological scenarii, our Universe experienced a dS phase in the very early epochs of its life \cite{Linde}.}, in part in the desire to establish possible mechanisms for \emph{late-time cosmology}\footnote{Recent astrophysical data coming from type Ia supernovae \cite{Riess} show that the expansion of our Universe is accelerating and point towards a small but nonvanishing positive cosmological constant. In other words, our Universe might presently be in a dS phase, which tends towards a pure dS spacetime.}, and in part in the need for a dS analogue of the so-called AdS/CFT correspondence (the dS/CFT correspondence). Yet, the underlying motivation behind this attempt stems from a more fundamental consideration/concern that we now elaborate on.

First, it is important to note that both the field theoretical reading and the phenomenological treatment of an elementary system, on the interpretation level in particular, rely heavily on the notions of energy, momentum, mass, and spin. The existence of such notions in turn is due to invariance principles - specifically the principle of invariance under the Poincar\'{e} group, that is, the group of motions of flat Minkowski spacetime. As such, in flat Minkowski spacetime, the rest mass $m$ and the spin $s$ of an (Einsteinian) elementary system are characterized by the two invariants that label the respective UIR of the Poincar\'{e} group (see Refs. \cite{Wigner, Newton/Wigner} and also footnote \ref{foot Wigner sense}).

In a curved spacetime, however, any interpretation of relativity with respect to the Poincar\'{e} group is physically irrelevant. Besides this, curved spacetimes in general (with the exception of the dS and AdS cases) admit no nontrivial groups of motion. It follows that there is no literal or unique way to extend the aforesaid physical notions to curved spacetimes as well (again, the dS and AdS are exceptions). As a matter of fact, while mathematically the important differential equations (Klein-Gordon and Dirac) can be easily generalized to forms that possess general covariance in curved spacetimes, the resulting formulations in the sense given in the previous paragraph cannot be interpreted in terms of physical elementary systems. To be frank, the modern theories of elementary systems are not primarily studies in differential equations \cite{Fronsdal}.

Nevertheless, in the sense given by Fronsdal \cite{Fronsdal}: \emph{``A physical theory that treats spacetime as Minkowskian flat must be obtainable as a well-defined limit of a more general physical theory, for which the assumption of flatness is not essential"}, the dS and AdS spacetimes are well-suited for generalizations of the notions of elementary systems. Actually, the dS and AdS spacetimes of respectively negative and positive constant curvatures constitute a particular family of curved spacetimes which, like their common null-curvature limit - flat Minkowski spacetime, admit continuous groups of motion of maximal symmetry; respectively, SO$_0(1,4)$ and SO$_0(2,3)$, or any of their covering groups. As such, the Poincar\'{e} group - the relativity group of flat Minkowski spacetime can be realized by a contraction limit of either the dS or the AdS relativity group.

\subsection{Methodology}
From now on, we particularly focus on the dS case which is of interest in the present study (for an overview of the AdS case, see Refs. \cite{Gazeau2022, Gazeau2023}).

On the representation level, UIRs of the dS group fall basically into three distinguished series, respectively known as principal, complementary, and discrete series \cite{Thomas, Newton, Takahashi, Dixmier, Martin1974}. The Poincar\'{e} massive UIRs can be totally realized by the null-curvature contraction limit of the dS principal series UIRs \cite{Mickelsson, Garidi}. In this sense, the latter are usually called dS (strictly) massive UIRs. Note that the \emph{spin}\footnote{The dS UIRs are generally labeled by two invariant parameters of the mass (energy) scale and the spin meaning. While the above arguments more or less give us a clue as to how the mass scale is raised in dS relativity, the existence of the spin notion in dS relativity needs some clarifications. To get the gist, we point out that the Poincar\'{e} contraction limit of the dS group is carried out with respect to the Lorentz subgroup SO$_0(1,3)$ (or its universal covering SL$(2, C)$). This technically means that the Lie algebra of the Lorentz subgroup, admitting the rotations algebra $\mathfrak{su}(2)$ as its closed subalgebra, remains unchanged during the contraction process. In other words, the dS group shares the same Lorentz subgroup SO$_0(1,3)$ (or SL$(2,C)$), and consequently, the same rotations subgroup SO$(3)$ (or its covering SU$(2)$), with the Poincar\'{e} one. This is actually the very point that gives sense to the notion of spin in dS relativity, strictly speaking, in those dS UIRs \emph{meaningful from the Minkowskian point of view}, because it stems from the same SO$(3)$ (or SU$(2)$) subgroup that the notion of spin in Einstein-Poincar\'{e} relativity does. For more details, see Ref. \cite{Gazeau2022}.}-$s$ fields associated with the dS (strictly) massive UIRs, admitting no gauge invariance, possess $2s + 1$ degrees of freedom. The situation for the realization of the massless UIRs of the Poincar\'{e} group through the dS UIRs is more subtle since among the dS UIRs there is no UIR comparable to the Poincar\'{e} massless infinite-spin UIRs. Nevertheless, a particular member of the dS scalar complementary series UIRs along with the dS higher-spin ($s>0$) discrete series UIRs lying at the lower end of this series form a set of the dS UIRs with a unique extension to the conformal group (SO$_0(2,4)$) UIRs in such a way that this extension is precisely equivalent to the conformal extension of the massless UIRs of the Poincar\'{e} group \cite{Barut, Mack}. This remarkable feature in a well-defined way allows us to distinguish the Poincar\'{e} massless UIRs with respect to the aforementioned set of the dS UIRs. In this sense, the representations belonging to the latter set are usually called the dS (strictly) massless UIRs. For the spin-$s\,(>0)$ fields associated with the dS (strictly) massless UIRs, due to the gauge-invariant properties, the degrees of freedom reduce to $2$, namely, the $2$ modes of helicities $\pm s$, while the propagation of these modes is confined to the light cone \cite{Flato1986, Angelopoulos1981}. On the other hand, the massless scalar field is commonly known as the conformally coupled scalar field.\footnote{It is important to note that, in the above sense, the so-called dS ``massless minimally coupled'' scalar field, associated with a specific scalar discrete series, is not actually a massless field, contrary to its name.} Note that all other dS UIRs (not being the (strictly) massive or massless ones) either have a nonphysical Poincar\'{e} contraction limit or do not have such a limit at all.

Here, we must underline that the last sentence given above by no means implies that the remaining part of the dS UIRs with no Minkowskian counterpart is physically irrelevant. On the contrary, it is perfectly legitimate to study all the dS UIRs in a consistent context, on both the mathematical (group representation) and the physical (field quantization) sides. For instance, among those dS UIRs with no Minkowskian interpretation, there is a particular class of UIRs that are of interest in the present study, known as partially massless UIRs. The latter are identified with those dS higher-spin ($s>3/2$) discrete series UIRs, which are not (strictly) massless, and are not ``contiguous" to the principal series \cite{Garidimass} (see also Ref. \cite{Gazeau2022}). The fields associated with these representations (say partially massless fields \cite{Deser1, Deser2, Deser3, Deser4, Deser5, Deser6, Higuchi}), due to a novel gauge invariance which was first noticed by Deser et al. in the case of spin-$2$ fields (see Ref. \cite{Deser1}), possess degrees of freedom less than the $2s + 1$ of their respective (strictly) massive fields and more than the $2$ of the (strictly) massless ones. Note that the naming convention ``partially massless" originates from the light-cone propagation properties of these fields \cite{Deser6}.

In this paper, we particularly study the dS partially massless spin-$2$ (say partially massless graviton) field and its covariant quantum field theory (QFT) reading in the above group theoretical context. We show that this field in the Wigner sense is associated with the discrete series UIR $\Pi^{\pm}_{2,1}$ (in the Dixmier notations \cite{Dixmier}). Strictly speaking, due to the combined occurrences of gauge invariance and indefinite metric, the complete, nondegenerate, and dS-invariant space of the respective quantum states is defined according to an indecomposable representation of the dS group admitting $\Pi^{\pm}_{2,1}$ as its central part. We employ the Wightman and G\"{a}rding axiomatic machinery for gauge invariant fields (Gupta-Bleuler scheme) to construct the respective QFT formulation \cite{Wightman}; the formalism of Gupta-Bleuler triplet underlies the indecomposable group representation structure and also allows for an explicit description of the gauge degree of freedom. However, as we now explain, this robust mathematical framework is not yet sufficiently completed to provide us with a consistent QFT formulation of the theory.

As soon as one steps on the path outlined above, one encounters the very problem of the absence of a true spectral condition, which, regardless of what machinery of QFT is employed, plagues QFT in dS spacetime. As a matter of fact, while generally in the context of dS QFT, it is rather straightforward to carry over the requirements of covariance and locality (microcausality) from the Minkowskian case, there is no literal or unique dS analogue of the ordinary spectral condition of ``positivity of the energy". To be more precise, there is no dS definition of the ``energy" concept at all. Technically, this critical problem is rooted in the absence of a canonical choice of a time coordinate in dS spacetime; no matter which dS group generator is considered, the corresponding Killing vector field, though perhaps timelike in some area of dS spacetime, is spacelike in some other area. In the absence of a true spectral condition - a canonical choice of time coordinate, based upon which one can distinguish positive and negative frequency modes (or in other words, distinguish a unique vacuum state), many inequivalent QFT formulations arise for any single dS field model. Therefore, besides the aforesaid symmetry considerations and the well-known Wightman and G\"{a}rding axioms, our QFT reading of elementary systems living in dS spacetime still needs to be supplemented by a criterion - the dS counterpart of the ordinary spectral condition.

Of course, here, one must notice that if attention is merely restricted to dS QFT on the free field level, there is a way out through the \emph{Hadamard condition} to get rid of the absence of a true spectral condition and to single out a specific vacuum state in the context of dS QFT. The Hadamard condition, which is actually a manifestation of the aforementioned Fronsdal principle, postulates that two-point functions of linear fields, for instance, Klein-Gordon fields, on a wide class of spacetimes with bifurcate Killing horizons, including dS spacetime, at short distances (in a tangent plane) should asymptotically meet their Minkowskian counterparts (see Refs. \cite{Haag, Allen, Kay} and references therein). The distinguished vacuum state for dS linear fields through the Hadamard condition exactly coincides with the vacuum state known in the literature under the name of Euclidean \cite{Gibbons} or Bunch-Davies \cite{Bunch}. Nevertheless, if one desires to step beyond the free-field level and deal with general interacting fields, the too-special character of the Hadamard condition (which makes it applicable only on the free-field level) compels one to look for another explanation of the existence of distinguished vacuum states in the dS global structure.

In $1990$s and in view of the above considerations, a remarkable elegant idea has been put forward by Bros et al. in their seminal works \cite{GazeauPRL, Bros 2point func}.\footnote{We also refer to Refs. \cite{B1, B2, B3, Neeb1, Neeb2, Massive/Massless 1/2, Massless 1, Massive 1, BehrooziTakook, Massive 2, Massive 3/2} and references therein for closely related discussions.} The idea, in a one-line description, lies in an appropriate adaptation of some familiar concepts of complex Minkowski spacetime to the complex dS one, strictly speaking, the adaptation of the very point that the analytic continuation properties of the QFT in the complexified Minkowski spacetime are directly linked to the energy content - specifically, to the spectral condition of the respective model. On this basis, the authors have argued that all ambiguities for dS QFTs in selecting preferred vacuum states can be lifted such that, despite the thermal features of the selected vacuum states (in the sense of the Gibbons-Hawking temperature \cite{Gibbons, Kay}), they exactly meet their Minkowskian counterparts at the null-curvature limit.\footnote{Note that, in this context, the importance of the Hadamard condition retrieves in another sense. The Hadamard condition guarantees that two-point functions are the boundary values of analytic functions ``from the good side" (the so-called $\mathrm{i}\epsilon$-rule.)} This adaptation is technically performed through a genuine, global dS-Fourier type calculus realized in terms of \emph{dS plane waves}. The latter, being introduced in their relevant tube domains, play an analogous role to plane waves in Minkowski spacetime. They are independent of any choice of coordinates, are well adapted to the representations of the dS group, and allow the yielded dS QFT to recover in a very suggestive way its flat Minkowskian counterpart under vanishing curvature (if such a limit exists!). In this context, Bros et al. \cite{GazeauPRL, Bros 2point func} have shown that, for instance, on the first level of complexity, namely, the free field level of dS QFTs, the usual spectral condition is replaced by a certain geometric Kubo-Martin-Schwinger (KMS) condition \cite{Kubo, Martin}, equivalent to a precise thermal characterization of the respective Euclidean/Bunch-Davies vacuum states.

Considering all the above, the dS group representation theory in the Wigner sense and the Wightman and G\"{a}rding axiomatic approach equipped with analyticity prerequisites in the complexified dS spacetime constitute the basis of our model in the QFT formulation of elementary systems - particularly, in this paper, the partially massless graviton living in dS spacetime.

Yet, a universal substitute for the concept of mass in dS relativity is a topic that is left open. To fulfill this legitimate demand, we take into account the Garidi mass formula \cite{Garidimass}. The latter, being defined in terms of the invariant parameters labeling the UIRs of the dS group, provides us with a consistent and univocal definition of the mass concept in dS relativity, which precisely gives sense to notions such as dS ``massive" and ``massless" fields according to their flat Minkowskian twins. It also has the advantage that it includes comprehensively all the mass formulas already introduced in the dS context (for more details, see Refs. \cite{Gazeau2022, Garidimass}).

\subsection{Limitations of analyticity-based quantization in dS fields}
It is crucial to emphasize that the quantization approach, based on the earlier-discussed analyticity requirement, encounters limitations when applied to specific dS fields. Notably, this limitation affects scalar fields linked to the scalar discrete series representations, as well as the dS graviton field associated with the discrete massless spin-2 representation. Each of these fields exhibits a distinctive form of gauge invariance that becomes anomalous at the quantum level. This gauge anomaly renders the theory inconsistent and necessitates resolution at all costs. A direct consequence of this inconsistency is the absence of assured dS-invariant Euclidean/Bunch-Davies vacuum states for these specific fields.

Addressing the aforementioned anomaly requires departing from the established framework and embracing an alternative approach grounded in a \emph{Krein structure (endowed with an indefinite inner product)}, rather than the conventional \emph{Hilbertian} one; the \emph{Krein QFT construction} versus the \emph{Hilbertian one}. This alternative approach remarkably guarantees that the theory possesses all the properties one might expect from a free field in dS spacetime with high symmetry, namely, the positivity of the norm of all physical states, adherence to causality, (full) covariance, and positivity of the energy operator in all physical states. [For a deeper grasp of these concepts, extending beyond the scope of this paper, we direct readers to references such as Refs. \cite{Massless 2, Massless 2', Massless 2'', Bamba 1, dS gravity 1, dS gravity 2, Gupta 2000, deBievre'} and also \cite{CCP-Krein, Casimir-Krein}.]

\textbf{\emph{Important remark:}} This framework bears the potential to offer new insights into comprehending the dS swampland conjecture \cite{Vafa1} from an alternative perspective distinct from the conventional Hilbertian approach. The essence of the dS swampland conjecture revolves around the idea that in a unified theory of quantum gravity, the viability of stable dS vacua, especially those pertinent to inflationary scenarios, could be restricted or even elusive.

\subsection{Layout}
To achieve our goal, the rest of this paper is organized as follows. In section \ref{Sec. II}, employing dS ambient space formalism, we present the dS machinery, geometry, group, and representation. In section \ref{Sec. III}, we introduce the corresponding gauge-invariant field equation in terms of the dS quadratic Casimir operator. We reveal the presence of the Gupta-Bleuler structure, which carries the indecomposable structure of the unitary representation of the dS group relevant to our problem. Additionally, we provide a general solution to the field equation along with its interpretation in terms of dS plane waves. In section \ref{Sec. IV}, we establish the corresponding Wightman two-point function ${\cal{W}}_{\alpha\beta\alpha^\prime\beta^\prime}(x,x^\prime)$ that satisfies the fundamental requirements of the field equation, namely, locality, covariance, and normal analyticity. The property of normal analyticity allows us to interpret ${\cal{W}}_{\alpha\beta\alpha^\prime\beta^\prime}(x,x^\prime)$ as the boundary value of an analytic two-point function $W_{\alpha\beta\alpha^\prime\beta^\prime}(z,z^\prime)$ from the (relevant) tube domains. The analytic kernel $W_{\alpha\beta\alpha^\prime\beta^\prime}(z,z^\prime)$ is defined in terms of dS waves within their respective tubular domains. Subsequently, we explicitly establish the Hilbert space structure and derive the field operator ${\cal{K}}(f)$. Furthermore, we provide a coordinate-independent formulation for the unsmeared field operator ${\cal{K}}(x)$. We finally summarize our results in section \ref{Sec. Conclusion}. This paper is supplemented with two appendices. Appendix \ref{App. Plane Waves} provides a brief introduction to dS plane waves, while Appendix \ref{App. bitensors} establishes a relationship between our construction and the maximally symmetric bitensors introduced in Ref. \cite{AllenJacobson}.

\subsection{Convention}
All over this article (unless noted otherwise!), we take into account the units $c = 1 = \hbar$, $c$ and $\hbar$ being the speed of light and the Planck constant, respectively.

\section{$\text{dS}$ geometry and relativity}\label{Sec. II}
In this section, following the lines sketched in Ref. \cite{Gazeau2022}, we briefly review a set of definitions, notations, and group-theoretical materials relevant to the description of the $1+3$-dimensional dS geometry and wave equations of elementary systems living on it.

\subsection{dS manifold and its causal structure}
Topologically, dS spacetime is $\mathbb{R}^1\times\mathbb{S}^3$, while $\mathbb{R}^1$ is a timelike direction. Geometrically, this spacetime is a globally hyperbolic spacetime of constant radius of curvature $R$. The dS hyperboloid can be conveniently described by its embedding in a $1+4$-dimensional Minkowski spacetime $\mathbb{R}^{5}$ (by the abusive identification $\mathbb{R}^{1+4} \equiv \mathbb{R}^{5}$) as:
\begin{eqnarray}\label{dS-M_R}
{M}_R \equiv \Big\{x = (x^0, \;...\; , x^4) \in\mathbb{R}^5 \;;\; \hspace{2cm}\nonumber\\
(x)^2 \equiv x\cdot x = \eta^{}_{\alpha\beta}x^\alpha x^\beta = -R^2 \Big\}\,,
\end{eqnarray}
where the indices $\alpha$ and $\beta$ take the values $0,1,2,3,4$, $x^\alpha$s are the respective Cartesian coordinates, and $\eta^{}_{\alpha\beta} = \mbox{diag}(1,-1,-1,-1,-1)$ is the ambient Minkowski metric. From a cosmological point of view, the constant radius $R$ may be identified with $R = H^{-1}$, where $H$, being the Hubble constant, gives the expansion rate of the spatial parts of dS spacetime.

The global causal ordering of dS spacetime is induced by that of ambient Minkowski spacetime; let:
\begin{eqnarray}\label{causal ordering dS2}
{V}^+ \equiv \Big\{x \in\mathbb{R}^5 \;;\; (x)^2 = x\cdot x \geqslant 0,\; x^0 > 0 \Big\}\,.
\end{eqnarray}
[For latter use, also let $\interior{V}^+ \equiv \big\{x \in\mathbb{R}^5 \;;\; (x)^2 > 0,\; x^0 > 0 \big\}$ denote the interior of ${V}^+$.] An ``event" $x^\prime \in {M}_R$ is said future connected to another one $x \in {M}_R$ (for the sake of simplicity, we symbolically write $x^\prime\geqslant x$), if $x^\prime - x \in {{V}^+}$, namely, $(x^\prime - x)^2 \geqslant 0$ (or equivalently,\footnote{Note that $(x^\prime - x)^2 = -2(R^2 + x\cdot x^\prime)$, for $x,x^\prime \in {M}_R$.} $x\cdot x^\prime \leqslant -R^2$), with $(x^{\prime 0} - x^0) > 0 $. In this sense, the future and past cones of an event $x \in {M}_R$, denoted respectively by ${\Sigma}^+(x)$ and ${\Sigma}^-(x)$, are:
\begin{eqnarray}
{\Sigma}^{+}(x) &=& \Big\{ x^\prime \in {M}_R \;;\; x^\prime \geqslant x \big)\Big\}\,,\nonumber\\
{\Sigma}^{-}(x) &=& \Big\{ x^\prime \in {M}_R \;;\; x^\prime \leqslant x \big)\Big\}\,,
\end{eqnarray}
and the respective ``light-cone" $\partial{\Sigma}(x)$, as the boundary set of ${\Sigma}^+(x) \bigcup {\Sigma}^-(x)$, is:
\begin{eqnarray}
\partial{\Sigma}(x) = \Big\{ x^\prime \in {M}_R \;;\; (x^\prime - x)^2 = 0 \hspace{2cm}\nonumber\\
\big(\mbox{or equivalently,}\; x\cdot x^\prime = -R^2\big) \Big\}\,.
\end{eqnarray}
Two events $x,x^\prime \in {M}_R$ are recognized as ``spacelike separated" or ``in acausal relation", if $x^\prime \notin {\Sigma}^+(x) \bigcup {\Sigma}^-(x)$, namely, if $(x^\prime - x)^2 < 0$ (or equivalently, $x\cdot x^\prime > -R^2$).

\subsection{Computation in ambient notations}
According to ambient space notations, dS fields are identified with symmetric (spinor-)tensor fields $\Psi^{(r)}_{\alpha_1 \,...\, \alpha_n}(x)$ on the dS manifold ($x \in M_R$), such that the indices $\alpha_1, ...\,, \alpha_n$ take the values $0,1,2,3,4$ and $r$ the values $n+1/2$, $n$ being the tensorial rank.\footnote{Here, for the sake of simplicity, we have ignored the spinorial index labeling the four spinorial components. From now on, we will also drop the tensorial indices, whenever it is possible, to simplify the notations.} These fields as functions of $\mathbb{R}^5$ are assumed to be homogeneous with some arbitrarily given degree of homogeneity $\ell$:
\begin{eqnarray}\label{homoge}
x\cdot\partial \Psi^{(r)}_{\alpha_1 \,...\, \alpha_n}(x) \; \Big( \equiv x^\alpha \frac{\partial}{\partial x^\alpha} \Big) = \ell \Psi^{(r)}_{\alpha_1 \,...\, \alpha_n}(x) \,,
\end{eqnarray}
where, for the sake of simplicity, the degree of homogeneity $\ell$ is usually set to $0$. [On this basis, for instance, by employing the identities that will be given in the section \ref{Subsubsec Link to intrinsic coordinates}, one can simply check that the d'Alembertian operator $\square_R \equiv \nabla_\mu \nabla^\mu$ defined on $M_R$ (where $\nabla_\mu$, with $\mu=0,1,2,3$, stands for the covariant derivative in local (intrinsic) coordinates) meets its twin $\square_5 \equiv \partial^2$ on $\mathbb{R}^5$.]

The fields are also assumed to be transitive with respect to all indices $\alpha_1, \,...\,, \alpha_n$:
\begin{eqnarray}\label{transversality}
x^{\alpha_i}_{} \Psi^{(r)}_{\alpha_1 \;...\; \alpha_i \;...\; \alpha_n}(x) = (x\cdot \Psi)^{(r-1)}_{\alpha_1 \;...\; \breve{\alpha}_i \;...\; \alpha_n}(x) = 0\,,
\end{eqnarray}
where by $\breve{\alpha}_i$ we mean this index is omitted. Clearly, the transversality requirement (concisely, say $x \cdot \Psi^{(r)}(x) = 0$) assures that the field $\Psi^{(r)}_{\alpha_1 \,...\, \alpha_n}(x)$ lies in the dS tangent spacetime.

Here, in view of the importance of the transversality requirement, the symmetric, ``transverse projector" $\theta^{}_{\alpha\beta} = \eta^{}_{\alpha\beta} + R^{-2} x_\alpha x_\beta$ ($\theta^{}_{\alpha\beta} x^\alpha = 0 = \theta^{}_{\alpha\beta} x^\beta$) is put forward; this operator is in fact the transverse form of the dS metric in ambient space formalism (we will clarify this important point in section \ref{Subsubsec Link to intrinsic coordinates}). Technically, $\theta^{}_{\alpha\beta}$ is used to construct transverse entities, like the transverse derivative $\overline{\partial}_\alpha = \theta^{}_{\alpha\beta}\partial^\beta = \partial_\alpha + R^{-2} x_\alpha x\cdot\partial$.\footnote{We note in passing that the transverse derivative $\overline{\partial}$ verifies the identities $\overline{\partial}_\alpha x_\beta = \theta^{}_{\alpha\beta}$ and $\overline{\partial}_\alpha (x)^2=0$. The latter reveals that $\overline{\partial}$ commutes with $(x)^2$, and hence, is intrinsically defined on the dS manifold $(x)^2 = - R^{2}$.} For a general (spinor-)tensor field $\Psi^{(r)}_{\alpha_1 \,...\, \alpha_n}(x)$, dedicating to each tensorial
index a specific transverse projector, as:
\begin{eqnarray}
\left( \prod_{i=1}^n \theta^{\beta_i}_{\alpha_i} \right) \Psi^{(r)}_{\beta_1 \,...\, \beta_n}(x) \equiv ({\cal{T}} \Psi)^{(r)}_{\alpha_1 \,...\, \alpha_n}(x)\,,
\end{eqnarray}
guarantees the transversality of the field in each tensorial index. Note that the degree of homogeneity of $\theta^{}_{\alpha\beta}$, with respect to the $\mathbb{R}^5$-variables $x^\alpha$, is zero, and hence, the above instruction does not alter the degree of homogeneity of the field.

\subsection{dS relativity group and its representations}
The dS relativity group SO$_0(1,4)$ (or its universal covering Sp$(2,2)$) is the ten-parameter group of all linear transformations in the ambient Minkowski spacetime $\mathbb{R}^5$, which leave invariant the quadratic form $(x)^2 = \eta^{}_{\alpha\beta}x^\alpha x^\beta$, enjoy the determinant unity, and finally preserve the direction of the ``time" variable $x^0$. A familiar realization of the associated Lie algebra is obtained by the linear span of the following (ten) Killing vectors:
\begin{eqnarray}\label{Killing dS}
K_{\alpha\beta} = x_\alpha \partial_\beta - x_\beta \partial_\alpha\,, \;\;\;\;\;\;\; K_{\alpha\beta} = - K_{\beta\alpha}\,.
\end{eqnarray}

On the representation (quantum) level, in the Hilbert space of the symmetric, \emph{square integrable}\footnote{With respect to some invariant inner product of Klein-Gordon type or else.}, (spinor-)tensors $\Psi^{(r)}_{\alpha_1 \,...\, \alpha_n}(x)$ on ${M}_R$, the Killing vectors $K_{\alpha\beta}$ are represented by (essentially) self-adjoint operators $L^{(r)}_{\alpha\beta} = M^{}_{\alpha\beta} + S^{(n)}_{\alpha\beta} + S^{(\frac{1}{2})}_{\alpha\beta}$, where the orbital part reads as:
\begin{eqnarray}
M_{\alpha\beta} = - \mathrm{i} (x_\alpha \partial_\beta - x_\beta \partial_\alpha ) = - \mathrm{i} (x_\alpha \overline\partial_\beta - x_\beta \overline\partial_\alpha )\,,
 \end{eqnarray}
the action of the spinorial part $S_{\alpha\beta}^{(n)}$ on the tensorial indices is:
\begin{eqnarray}
S_{\alpha\beta}^{(n)}\Psi^{(r)}_{\alpha_1 \,...\, \alpha_n} &=& - \mathrm{i} \sum_{i=1}^{n}\Big(\eta^{}_{\alpha\alpha_i}\Psi^{(r)}_{\alpha_1 \,...\, (\alpha_i\mapsto \beta) \,...\, \alpha_n} \hspace{1cm} \nonumber\\
&& \hspace{2cm} - (\alpha \rightleftharpoons \beta)\Big)\,,
\end{eqnarray}
and finally the spinorial part $S^{(\frac{1}{2})}_{\alpha\beta}$ acts on the spinorial indices by $S^{(\frac{1}{2})}_{\alpha\beta} = -\frac{\mathrm{i}}{4}[\gamma_\alpha,\gamma_\beta]$, where $\gamma_\alpha$s stand for the five $4\times 4$-matrices generating the Clifford algebra (see Ref. \cite{Gazeau2022}). The self-adjoint operators $L^{(r)}_{\alpha\beta}$ obey the standard commutation relations of the dS Lie algebra:
\begin{eqnarray}\label{adjoint commutator}
\left[L^{(r)}_{\alpha\beta},L^{(r)}_{\gamma\delta}\right] \hspace{6cm}\nonumber\\
= - \mathrm{i} \left( \eta^{}_{\alpha\gamma} {L^{(r)}_{\beta\delta}} + \eta^{}_{\beta\delta} {L^{(r)}_{\alpha\gamma}} - \eta^{}_{\alpha\delta} {L^{(r)}_{\beta\gamma}} - \eta^{}_{\beta\gamma} {L^{(r)}_{\alpha\delta}} \right)\,.\nonumber
\end{eqnarray}
Note that the operators $L^{(r)}_{\alpha\beta}$ are intrinsically defined on the dS hyperboloid $(x)^2 = -R^{2}$, since we have $[L^{(r)}_{\alpha\beta},(x)^2]=0$.

In this group theoretical construction, there are two Casimir operators:
\begin{eqnarray}\label{Casimir 2}
\mbox{quadratic}\;&;&\;\; Q^{(1)}_r = - \frac{1}{2} L^{(r)}_{\alpha\beta} L^{(r)\alpha\beta}\,, \nonumber\\
\mbox{quartic}\;&;&\;\; Q^{(2)}_r = - W^{(r)}_\alpha W^{(r)\alpha}\,,
\end{eqnarray}
where the dS counterpart of the Pauli-Lubanski operator $W^{(r)}_\alpha = - \frac{1}{8} {\cal{E}}_{\tiny{\alpha\beta\delta\rho\sigma}} L^{(r)\beta\delta} L^{(r)\rho\sigma}$, in which ${\cal{E}}_{\tiny{\alpha\beta\delta\rho\sigma}}$ refers to the five-dimensional totally antisymmetric Levi-Civita symbol. These two Casimir operators are also intrinsically defined on the dS hyperboloid $(x)^2 = -R^{2}$ as $L^{(r)}_{\alpha\beta}$s do, since $[Q^{(1,2)}_r,(x)^2]=0$. Moreover, they commute with all generator representatives $L^{(r)}_{\alpha\beta}$, and hence, act like constants on all states in a certain dS UIR:
\begin{eqnarray}
Q^{(1,2)}_r \Psi^{(r)} = \langle Q^{(1,2)}_r \rangle \Psi^{(r)},
\end{eqnarray}
where the respective eigenvalues $\langle Q^{(1,2)}_r \rangle$, in the Dixmier notations \cite{Dixmier}, are determined in terms of a pair of parameters $\Delta(p,q)$, with $p\in \mathbb{N}/2$ and $q\in\mathbb{C}$, as:
\begin{eqnarray}
\label{Casimir rank 2}\langle Q^{(1)}_r \rangle &=& \big(-p(p+1) - (q+1)(q-2)\big)\,, \\
\label{Casimir rank 4}\langle Q^{(2)}_r \rangle &=& \big(-p(p+1)q(q-1)\big)\,.
\end{eqnarray}
Therefore, the spectral values assumed by the Casimir operators (say, the allowed values of $p$ and $q$) can be utilized to classify UIRs of the dS group. These UIRs, as already pointed out, fall basically into three distinguished series as we explain below (see Refs. \cite{Dixmier, Takahashi}).

\subsubsection{Principal series representations}
Principal series representations ${U}^{\mbox{\small{ps}}}_{s,\nu}$ are characterized by $\Delta(p=s,q=\frac{1}{2}+i\nu)$, where the parameter $p=s$ possesses a spin meaning. Here, two different cases must be distinguished:
\begin{itemize}
    \item{The integer spin representations, with $\nu\in \mathbb{R}$ and $s=0,1,2,...$ .}
    \item{The half-integer spin representations, with $\nu\in \mathbb{R}-\{0\}$ and $s=\frac{1}{2},\frac{3}{2},\frac{5}{2},...$ .}
\end{itemize}

Let us incorporate the parameters $c$ (the speed of light) and $\hbar$ (the Planck constant) without setting them to unity. In this context, we take from Ref. \cite{Gazeau2022} the relation between the representation parameter $\nu$ and the Poincar\'{e}-Minkowski mass $m$:
\begin{align}
\nu = \frac{mcR}{\hbar} - \left( s-\frac{1}{2} \right)^2 \left( \frac{\hbar}{2mcR} + {\mathcal{O}}\left( \frac{1}{c^2R^2}\right)\right)\,.
\end{align}
This approximation formula will be given its full meaning in section \ref{Subsec. Garidi} below. Then, the Poincar\'{e} contraction limit of the above representations (denoted below by `$\longrightarrow$') is technically performed when $\nu$ and $R$ tend to infinity (while $m = {\hbar\nu}/{cR}$ remains intact) \cite{Mickelsson, Garidi}:
\begin{eqnarray}\label{massive contraction}
{U}^{\mbox{\small{ps}}}_{s,\nu} \;\; \underset{\underset{\hbar \nu/cR = m}{R \rightarrow \infty,\; \nu \rightarrow \infty}}{\longrightarrow} \;\; {\cal{P}}^>_{s,m} \; \oplus \; {\cal{P}}^<_{s,m} \,,
\end{eqnarray}
where ${\cal{P}}^\gtrless_{s,m}$ respectively refer to the positive/negative energy Wigner UIRs of the Poincar\'{e} group, with spin $s$ and mass $m$. As already mentioned, the dS principal UIRs are recognized as dS massive representations, in the above sense.

Note that the above breaking of the irreducibility of the dS principal (massive) UIRs into a direct sum of two Poincar\'{e} massive ones possessing positive and negative energies can be cured either by taking into account a proper choice of the dS (global) modes (say dS plane waves), being properly defined in their (relevant) analyticity tube domains \cite{Garidi}, or by considering the Poincar\'{e} contraction of the representations in terms of a causality dS semi-group \cite{Mizony1984}. Then, we have:
\begin{eqnarray}\label{massive contraction'}
{U}^{\mbox{\small{ps}}}_{s,\nu} \;\; \underset{\underset{\hbar \nu/cR = m}{R \rightarrow \infty,\; \nu \rightarrow \infty}}{\longrightarrow} \;\; {\cal{P}}^>_{s,m}.
\end{eqnarray}

\subsubsection{Complementary series representations}
Complementary series representations ${U}^{\mbox{\small{cs}}}_{s,\nu}$ are characterized by $\Delta(p=s,q=\frac{1}{2}+\nu)$, where, again, the parameter $p=s$ has a spin meaning. Here also, two distinguished cases come to the fore:
\begin{itemize}
    \item{The scalar representation ${U}^{\mbox{\small{cs}}}_{0,\nu}$, with $\nu\in \mathbb{R}$ and $0<|\nu|<\frac{3}{2}$.}
    \item{The spinorial representation ${U}^{\mbox{\small{cs}}}_{s,\nu}$, with $\nu\in \mathbb{R}$ and $0<|\nu|<\frac{1}{2}$, while $s=1,2,3,...$ .}
\end{itemize}

The only \emph{meaningful} representation of the complementary series UIRs \emph{from the point of view of a Minkowskian observer} is the scalar representation ${U}^{\mbox{\small{cs}}}_{s=0, \nu=\frac{1}{2}}$ ($\Delta(p=0,q=1)$); it has a unique extension (denoted below by `$\hookrightarrow$') to the UIR ${\cal{C}}^{>}_{1,0,0}$ of the conformal group $\mathrm{SO}_0(2,4)$,\footnote{Note that conformal invariance technically entails the discrete series representations (and their lower end) of the (universal covering of the) conformal group or its double covering group $\mathrm{SO}_0(2,4)$ or its fourth covering group $\mathrm{SU}(2,2)$. Here, the respective conformal UIRs are characterized by ${\mathcal C}^{\gtrless}_{E_0,j_l, j_r}$, with the parameter $E_0$ denoting the positive/negative conformal energy and $(j_l,j_r) \in \mathbb{N}/2 \times \mathbb{N}/2$ labeling the UIRs of $\mathrm{SU}(2) \times \mathrm{SU}(2)$.} while this extension is equivalent to the conformal extension of the Poincar\'{e} massless scalar UIRs ${\cal{P}}^{\gtrless}_{0,0}$ (respectively, with positive/negative energy) \cite{Barut, Mack}:
\begin{eqnarray}
\left. \begin{array}{ccccccc}
& & {\cal{C}}^{>}_{1,0,0} & & {\cal{C}}^{>}_{1,0,0} & \hookleftarrow & {\cal{P}}^{>}_{0,0} \\
{U}^{\mbox{\small{cs}}}_{0,\frac{1}{2}} & \hookrightarrow & \oplus & \underset{R\rightarrow \infty}{\longrightarrow} & \oplus & &\oplus \\
& & {\cal{C}}^{<}_{-1,0,0} & & {\cal{C}}^{<}_{-1,0,0} & \hookleftarrow & {\cal{P}}^{<}_{0,0} \,.
\end{array} \right.
\end{eqnarray}
In the above sense, the scalar representation ${U}^{\mbox{\small{cs}}}_{0, \frac{1}{2}}$ is called massless.

\subsubsection{Discrete series representations}
Discrete series representations $\Pi_{p,q}^\pm$ are characterized by $\Delta(p,q)$, where, for the symmetric cases $\Pi^{\pm}_{p=s,q=s}$, the parameter $p=s$ (with $s>0$) has a spin (helicity) meaning; as a matter of fact, the superscript `$\pm$' stands for the helicities $\pm s$. Here, we have to distinguish between:
\begin{itemize}
    \item{The nonsquare-integrable scalar case $\Pi_{p,0}$, with $p=1,2,...$ .}
    \item{The spinorial cases $\Pi^{\pm}_{p,q}$, with $p = \frac{1}{2},1,\frac{3}{2},...$ and $q = p,p-1,...,1$ or $\frac{1}{2}$ ($q>0$); the representations characterized by $q = \frac{1}{2}$, namely, $\Pi^{\pm}_{p,\frac{1}{2}}$, are not square integrable.}
\end{itemize}

In the case of the discrete series UIRs, the meaningful representations from the Minkowskian point of view are the aforementioned symmetric cases $\Pi^{\pm}_{p=s,q=s}$ (with $s>0$) lying at the lower limit of this series; they have a unique extension to the conformal group UIRs, while this extension is equivalent to the conformal extension of the massless spin ($s>0$) UIRs of the Poincar\'{e} group \cite{Barut, Mack}:
\begin{eqnarray}
\left. \begin{array}{ccccccc}
& & {\cal{C}}^{>}_{s+1,0,s} & & {\cal{C}}^{>}_{s+1,0,s} & \hookleftarrow & {\cal{P}}^{>}_{-s,0} \\
\Pi^+_{s,s} & \hookrightarrow & \oplus & \underset{R\rightarrow \infty}{\longrightarrow} & \oplus & & \oplus \\
& & {\cal{C}}^{<}_{-s-1,0,s} & & {\cal{C}}^{<}_{-s-1,0,s} & \hookleftarrow & {\cal{P}}^{<}_{-s,0} \,,
\end{array} \right.
\end{eqnarray}
\begin{eqnarray}
\left. \begin{array}{ccccccc}
& & {\cal{C}}^{>}_{s+1,s,0} & & {\cal{C}}^{>}_{s+1,s,0} & \hookleftarrow & {\cal{P}}^{>}_{s,0} \\
\Pi^-_{s,s} & \hookrightarrow & \oplus & \underset{R\rightarrow \infty}{\longrightarrow} & \oplus & & \oplus \\
& & {\cal{C}}^{<}_{-s-1,s,0} & & {\cal{C}}^{<}_{-s-1,s,0} & \hookleftarrow & {\cal{P}}^{<}_{s,0}\,.
\end{array} \right.
\end{eqnarray}
In the above sense, the UIRs $\Pi^{\pm}_{s,s}$ (with $s>0$) are called massless.

\textbf{\emph{Important remark:}} For all three series of the dS UIRs, the respective Casimir eigenvalues do not alter by letting $q\mapsto (1-q)$. In other words, the representations given by the pairs $\Delta(p,q)$ and $\Delta(p,1-q)$ share the same Casimir eigenvalues. By definition, such representations are recognized as Weyl equivalent representations.

\subsection{dS field equations}\label{Subsec. dS field equ.}
As briefly pointed out in the introduction section, (projective) Hilbert spaces carrying the dS UIRs (in some restricted sense) identify the quantum (``one-particle") state spaces of the respective elementary systems living in dS spacetime. Such (projective) Hilbert spaces are densely generated by the \emph{square integrable}\footnote{Again, with respect to some invariant inner product of Klein-Gordon type or else (in our case, for instance, with respect to the inner product (\ref{inner product gen.})).}, (spinor-)tensors $\Psi^{(r)}_{\alpha_1 \,...\, \alpha_n}(x)$ on ${M}_R$. In practice, for a given dS UIR, the common dense subspace (of the respective Hilbert space) carrying the UIR is generated by the eigenfunctions of the dS Casimir operators $Q^{(1,2)}_r$ for the assumed eigenvalues $\langle Q^{(1,2)}_r \rangle$, namely:
\begin{eqnarray}
\left( Q^{(1,2)}_r - \langle Q^{(1,2)}_r \rangle \right) \Psi^{(r)}_{\alpha_1 \,...\, \alpha_n}(x) =0\,.
\end{eqnarray}
The corresponding ``wave (field) equation" is then identified with that of the quadratic Casimir operator $Q^{(1)}_r$:\footnote{The equation involving the quartic Casimir operator $Q^{(2)}_r$, possessing higher derivatives, naturally entails ``ghost" solutions.}
\begin{eqnarray}\label{Field Eq. Gen}
\left( Q^{(1)}_r - \langle Q^{(1)}_r \rangle \right) \Psi^{(r)}_{\alpha_1 \,...\, \alpha_n}(x) =0\,.
\end{eqnarray}

Now, for later use, let us give the explicit form of the field equation associated with a rank-$2$ tensor field, say, $\Psi^{(2)}_{\alpha\beta} \equiv {\cal{K}}_{\alpha\beta}$, in terms of the ambient space notations. The explicit form of the quadratic Casimir operator $Q^{(1)}_2$, acting on the space generated by the rank-$2$ tensors ${\cal{K}}_{\alpha\beta}$, reads as \cite{Gazeau2022}:
\begin{eqnarray}\label{Q1}
Q_2^{(1)} \; {\cal{K}}_{\alpha\beta} &=& \left( Q_0^{(1)} - 6 \right) {\cal{K}}_{\alpha\beta} - 2 {\cal{S}} \partial x\cdot {\cal{K}}_{\alpha\beta} \nonumber\\
&& + 2 {\cal{S}} x \partial\cdot {\cal{K}}_{\alpha\beta} + 2 \eta_{\alpha\beta} {\cal{K}}^\prime \,,
\end{eqnarray}
where $Q_0^{(1)}= -\frac{1}{2} M_{\alpha\beta}M^{\alpha\beta}= -R^{2} \overline\partial^2$, the symmetrizer operator ${\cal S}$ acts as ${\cal S}(\zeta_\alpha \omega_\beta)=\zeta_\alpha \omega_\beta + \zeta_\beta \omega_\alpha$, and finally ${\cal{K}}^\prime \equiv \eta^{\alpha\beta} {\cal{K}}_{\alpha\beta}$ denotes the trace of the field. Now, in the ambient space notations, the explicit form of the respective field equation can be achieved by substituting Eq. (\ref{Q1}) into (\ref{Field Eq. Gen}). But, once one proceeds with this substitution, due to the form of $Q_2^{(1)}$ in the ambient notations (given above), one encounters some invariant subspaces in the space of solutions to the field equation, which must be eliminated if one desires to be left with the space that merely carries the respective dS UIR. Accordingly, the requirements of homogeneity (reading here as $x \cdot \partial {\cal{K}} = 0$) and transversality ($x \cdot {\cal{K}} = 0$) intrinsic to a field in the ambient space notations must be supplemented by the divergenceless requirement ($\partial \cdot {\cal{K}} = 0$); note that the transversality and divergenceless requirements together entail ${\cal{K}}^{\prime} = 0$. The explicit form of the field equation for a rank-$2$ tensor field ${\cal{K}}_{\alpha\beta}$ then reads:
\begin{align}\label{Field Eq. Gen2}
&\left( Q^{(1)}_2 + p(p+1) + (q+1)(q-2) \right) {\cal{K}}_{\alpha\beta} \nonumber\\
&\quad\quad = \left( Q_0^{(1)} - 6 + p(p+1) + (q+1)(q-2) \right) {\cal{K}}_{\alpha\beta} = 0\,,
\end{align}
with the particular permissible values that the parameters $p$ and $q$ can take for the three series of dS UIRs. Of course, we will show in the sequel that the above field equation, in the presence of the gauge invariance property of the dS partially massless graviton field, is too restrictive and just determines the physical solutions. It actually needs to be modified in such a way that it involves the gauge solutions as well.

We would also like to provide the explicit expression of the field equation governing a scalar field $\Psi^{(0)}_{} \equiv \phi$. Two cases arise for scalar fields: the scalar principal and complementary series determined by $p=0$, and the scalar discrete series given by $q=0$. The field equation given below holds for both cases:
\begin{eqnarray}\label{Wave Eq. scalar}
\Big( Q^{(1)}_0 + \tau(\tau+3) \Big) \phi(x) = 0\,,
\end{eqnarray}
where the scalar principal series is characterized by the unifying complex parameter $\tau$, which takes the values $\tau = -q-1 = -3/2- \mathrm{i} \nu$ with $\nu \in \mathbb{R}$, the scalar complementary series by the values $\tau = -q-1 = -3/2-\nu$ with $\nu \in \mathbb{R}$ and $0 < |\nu| <3/2$, and finally the scalar discrete series by the values $\tau = p-1$ or $ \tau= -p-2$, with $p=1,2,...$ . Note that, according to Ref. \cite{Gazeau2022}, the solution to the given field equation is only valid for values of $\tau$ where $\mbox{Re}(\tau) < 0$. Therefore, for the scalar discrete series, we must exclude the values of $\tau = p-1$.

Note that, from now on, we simplify our notations by dropping the superscript `$(1)$' from $Q^{(1)}_r$ ($Q^{}_r \equiv Q^{(1)}_r$).

\subsubsection{Link to intrinsic coordinates}\label{Subsubsec Link to intrinsic coordinates}
Here, it would be useful to provide the link between the intrinsic and ambient coordinates. For the sake of simplicity, we just bring the relations which are relevant to the context of the present study, i.e., to a rank-$2$ tensor field and its field equation; similar relations hold for other (spinor-)tensor fields.

The corresponding intrinsic field $h_{\mu\nu}(X)$ is locally characterized by ${\cal{K}}_{\alpha\beta}(x)$ as:
\begin{eqnarray}\label{tt}
h_{\mu\nu}(X) = x^{\alpha}_{\,\,,\,\mu} x^{\beta}_{\,\,,\,\nu} \; {\cal{K}}_{\alpha\beta} \big(x(X)\big)\,,
\end{eqnarray}
where $x^{\alpha}_{\,\,,\,\mu} = \partial x^{\alpha}/\partial X^{\mu}$ and $X^\mu$s, with $\mu=0,1,2,3$, stand for the four local spacetime coordinates on $M_R$. The respective dS metric is achieved by inducing the natural ambient Minkowski ($\mathbb{R}^5$) metric on $M_R$:
\begin{eqnarray}
\mathrm{d} s^2= \eta_{\alpha\beta}\mathrm{d} x^{\alpha}\mathrm{d} x^{\beta}\big|_{(x)^2=-R^{2}} = g_{\mu\nu} \mathrm{d} X^{\mu}\mathrm{d} X^{\nu}\,.
\end{eqnarray}
Note that the only symmetric and transverse tensor which is linked to the dS metric, in view of Eq. (\ref{tt}), is $\theta^{}_{\alpha\beta}$; $g_{\mu\nu}= x^\alpha_{\,\,,\,\mu} x^\beta_{\,\,,\,\nu} \theta^{}_{\alpha\beta}$. The transformation of the covariant derivatives is also given by:
\begin{eqnarray}
\nabla_\rho \nabla_\lambda h_{\mu\nu} = x_{\,\,,\,\rho}^{\gamma}x_{\,\,,\,\lambda}^{\sigma} \;\; x_{\,\,,\,\mu}^{\alpha}x_{\,\,,\,\nu}^{\beta} \;\; ({\cal{T}}\overline\partial_\gamma) ({\cal{T}}\overline\partial_\sigma) \; {\cal{K}}_{\alpha\beta}\,.
\end{eqnarray}
Then, the d'Alembertian operator reads:
\begin{eqnarray}
\square_R \phi = g^{\mu\nu} \nabla_\mu \nabla_\nu \phi &=& g^{\mu\nu} x^\alpha_{\,\,,\,\mu} x^\beta_{\,\,,\,\nu} \Big( \overline\partial_\alpha \overline\partial_\beta - R^{-2} x^{}_\beta \overline\partial_\alpha \Big)\phi \nonumber\\
&=& \theta^{\alpha\beta} \Big( \overline\partial_\alpha \overline\partial_\beta - R^{-2} x^{}_\beta \overline\partial_\alpha \Big)\phi = \overline\partial^2 \; \phi\,,\nonumber
\end{eqnarray}
where $\phi$ is a scalar field in dS spacetime. Note that, considering the definitions subsequent to Eq. (\ref{Q1}), we have $Q_0 = - R^{2} \square_R$.

\subsection{Garidi mass and energy at rest in dS}\label{Subsec. Garidi}
Let, once again, the speed of light $c$ and the Planck constant $\hbar$ be no longer normalized to unity.

We here adopt the definition of mass in dS relativity proposed by Garidi \cite{Garidimass}:
\begin{eqnarray}\label{conmassf}
\mathfrak{M}^2 = \frac{\hbar^2}{c^2R^2}\Big( \langle Q^{}_r\rangle - \langle Q_r^{p=q=s}\rangle \Big)\,,
\end{eqnarray}
where: (i) the dS radius of curvature is given by $R = \sqrt{3/\Lambda} = c/H$, where $\Lambda$ represents the (positive) cosmological constant and $H$ the Hubble constant, (ii) $\langle Q^{}_r\rangle$ refers to the eigenvalues \eqref{Casimir rank 2} of the dS quadratic Casimir operator corresponding to the dS UIR indexed by the pair $\Delta(p,q)$, (iii) $\langle Q_r^{p=q=s}\rangle$ refers to the eigenvalues corresponding to the dS discrete (massless) UIRs $\Pi^\pm_{s,s}$:
\begin{eqnarray}\label{eigdisse}
\langle Q_r^{p=q=s}\rangle = -2(s^2-1)\,,
\end{eqnarray}
which, for $s=0$, is also equal to the value $2$ of the Casimir operator for the conformally massless scalar UIR in the complementary series. 

Accordingly, the Garidi mass is zero for all (conformally) massless dS UIRs, and is nonzero for all other UIRs. Precisely, from the point of view of a Minkowskian observer, the Garidi mass definition \eqref{conmassf} is meaningful for:
\begin{itemize}
\item{the entire set of principal (massive) UIRs ${U}^{\mbox{\small{ps}}}_{s,\nu}$, with $s\in\mathbb{N}/2$ and $\nu\in\mathbb{R}$,}
\item{the scalar complementary (massless) UIR ${U}^{\mbox{\small{cs}}}_{0,\frac{1}{2}}$,}
\item{and finally, the discrete (massless) UIRs $\Pi^\pm_{s,s}$, with $s>0$, lying at the lower limit of this series.}
\end{itemize}
The fact that the lowest values of $\langle Q^{}_r\rangle$ occur at $p=q=s$ guarantees the non-negativity of the Garidi mass definition ($\mathfrak{M}^2\geqslant 0$) for such representations. Moreover, for a given dS principal (massive) UIR ${U}^{\mbox{\small{ps}}}_{s,\nu}$, the (positive) Garidi mass $\mathfrak{M}$ is given by
\begin{equation}
\label{garidips}
\mathfrak{M}=\frac{\hbar}{Rc}\left[\nu^2 + \left(s-\frac{1}{2}\right)^2\right]^{1/2}\, ,
\end{equation}
and, in the null-curvature limit, consistently  converges to the Minkowski mass $m$ of the respective Poincar\'{e} massive UIR.

In the case of a dS UIR with no meaningful Minkowskian interpretation, it is important to note that the Garidi mass formula (\ref{conmassf}) can still be applied, but without invoking a Minkowskian interpretation. Particularly, in the case of the complementary series, (\ref{garidips}) explicitly reads as:
\begin{equation}
\label{garidics}
\mathfrak{M}=\frac{\hbar}{Rc}\left[\left(s-\frac{1}{2}\right)^2-\nu^2\right]^{1/2}\, .
\end{equation}
This definition is valid within the interval $0<\vert\nu\vert<\vert s-\frac{1}{2}\vert$. Consequently, it does not extend to cover all scalar complementary representations (with $0<|\nu|<\frac{3}{2}$). Instead, its validity is confined to the range of $0<\vert\nu\vert< \frac{1}{2}$. Furthermore, it is evident that this definition does not account for the scenario where $s=\frac{1}{2}$.

In the case of the dS partially massless graviton field, which is of particular interest in the present study, and which (in a shortcut) corresponds to the discrete UIR $\Pi^{\pm}_{2,1}$ ($\Delta(p=2,q=1)$), the Garidi mass formula yields $\mathfrak{M}^2 = 2H^2 = 2\Lambda/3$. This outcome coincides precisely with the result previously reported by others \cite{Deser1, Deser2, Deser3, Deser4, Deser5, Deser6, Higuchi}. Again, it is worth highlighting the advantage of the Garidi mass formula, as it comprehensively encompasses all the previously introduced mass formulas within the dS context.
For a comprehensive review of various mathematical/physical aspects of the Garidi mass formula (\ref{conmassf}), readers are referred to Ref. \cite{Gazeau2022}.

The Garidi mass $\mathfrak{M}$ within dS spacetime is anticipated to remain inherently invariant, regardless of the surrounding spacetime geometry. This leads to the absence of distinction between inertial and gravitational mass within the classical theory framework, reflecting a direct embodiment of the equivalence principle. Since for dS massive representations $\mathfrak{M}\to_{\Lambda \to 0} m$ and for dS  massless representations $\mathfrak{M}=0$ the natural expectation arises that $\mathfrak{M}$ is the proper mass $m$ of the considered elementary system.

In the framework of Minkowski spacetime, the rest energy ($E^{\text{rest}}$) of an elementary system precisely corresponds to its proper mass ($m$) (up to the $c^2$ factor). However, this alignment unravels within the domain of dS spacetime \cite{Gazeau2022, Garidimass, Tannoudji}. Specifically, the concept of rest energy ($\mathfrak{E}^\text{rest}$) within dS space encounters intricacy due to the inherent uncertainty arising from the definition of time. However, in the case of the principal series with $\nu\geqslant 0$ a consistent definition can be established from \eqref{garidips} as:
\begin{equation}
\label{dsErest}
{\mathfrak{E}}^{\text{rest}} \equiv \frac{\hbar c\nu}{R} =\left[m^2c^4 -\frac{ \hbar^2 c^2 }{R^2}\left(s - \frac{1}{2}\right)^2\right]^{1/2}
\end{equation} 
in dS relativity \cite{Gazeau2022}.
The two  concepts of energy at rest and proper mass, which are distinct for $s\neq \frac{1}{2}$, merge at the limit, as one should expect. 

Moreover, for all cases where the Garidi mass  cannot be identified with a ``universal'' proper mass, for instance, as it is the case for the partially massless fields studied  in this paper, it is still allowed to consider the invariant quantity $\mathfrak{M}c^2$ as a kind of ``energy at rest'', being equal to a ``dS proper mass''.

\subsection{A brief discussion: Lifetime of a dS particle}\label{Subsec. Lifetime}
Referring to Ref. \cite{Lifetime}, it is worth noting a substantial observation within the discussed group theoretical construction. Interestingly, this observation points towards a potential connection with the distance conjecture within the context of string theory \cite{Vafa2}.

In Ref. \cite{Lifetime}, the authors have focused their investigation on the dS scalar representations. They have introduced an alternate definition of mass for the examined representations, distinct from the Garidi mass. This particular mass parameter is just proportional to $\nu\geqslant 0$ and is denoted here as $M$ to differentiate it from the Garidi mass $\mathfrak{M}$. In this context, they have demonstrated that the tensor product of two principal UIRs with masses ${M}_1$ and ${M}_2$ undergoes decomposition into a direct integral of representations whose masses ${M}$ do not adhere to the ``sub-additivity condition'' ${M} \geqslant {M}_1 + {M}_2$. This observation underscores that the dS symmetry does not hinder a particle with mass in the principal series from undergoing decay, leading to the creation of heavier particle pairs as an example; this phenomenon implies the absence of a mass gap within this range. Interestingly, the lifetime of such a particle remains unchanged regardless of its velocity, especially when that lifetime is comparable to the dS radius.

The authors have also demonstrated that, in contrast, the tensor product of two complementary representations encompasses an additional finite sum of discrete terms within the complementary series itself (with at most one term in dimension four). This underscores a form of particle stability. The novelty here is that a particle of this nature cannot undergo disintegration unless the masses of the resultant decay products possess specific quantized values. However, as it is apparent from the relation \eqref{garidics}, one should be aware that $\mathfrak{M}$ decreases from $\frac{1}{2}$ to $0$ as $\vert\nu\vert$ increases from $0$ to $\frac{1}{2}$.

For a comprehensive exploration within this context, extending beyond the confines of the present study, we direct readers to Ref. \cite{Lifetime} and the references cited therein.

\section{Partially massless graviton field equation and the space of solutions}\label{Sec. III}

\subsection{Field equation}

\subsubsection{Gauge-invariant field equation}
We are now in a position to study the partially massless spin-$2$ (say partially massless graviton) field in the context of dS relativity. As we will show in the sequel, this field is associated with the discrete series UIR $\Pi^{\pm}_{2,1}$, characterized by $\Delta(p=2,q=1)$, for which we have $\langle Q_2 \rangle=-4$. Therefore, the corresponding field equation should be as follows:
\begin{eqnarray}\label{Field Eq. Gen2'}
\left( Q_2 +4 \right) {\cal{K}} = 0\,,
\end{eqnarray}
while $x \cdot \partial {\cal{K}} = 0$, $x \cdot {\cal{K}} = 0$, and $\partial\cdot {\cal{K}}=\overline\partial\cdot {\cal{K}}=0$\footnote{Note that generally, for a transitive ${\cal{K}}$, we have $\partial\cdot {\cal{K}}=\overline\partial\cdot {\cal{K}}$.} (consequently, ${\cal{K}}^{\prime} = 0$).

To establish a general solution to the above field equation, we begin with producing a recurrence formula that presents the rank-$2$ tensor field ${\cal{K}}_{\alpha\beta}$ in terms of lower ranks tensors (scalar and vectors). This process technically involves operators that are expected to obey commutation/intertwining rules with $L_{\alpha\beta}^{(2)}$s and $Q_2$. The contraction of the transverse projector $\theta$ with a constant polarization five-vector $Z$, namely, ${\cal S} \theta\cdot Z \equiv {\cal S}\overline{Z}$, is, for instance, an ingredient part of such a recurrence formula since it allows for defining an operator which makes a symmetric transverse rank-$2$ tensor field ${\cal K}$ from a transverse rank-$1$ tensor field $\widetilde{K}$. The following commutation relation holds between $Q_2$ and ${\cal S} \overline{Z} \widetilde{K}$:
\begin{align}\label{recur1}
Q_2{\cal S}\overline{Z}\widetilde{K} = {\cal S}\overline{Z}(Q_1-4)\widetilde{K} -2R^{-2}{D_2} (Z\cdot x)\widetilde{K} +4{\theta} Z\cdot \widetilde{K}\,,
\end{align}
where $D_2= {\cal S}(D_1 - x)$, with $D_1= R^{2}\overline\partial$. [Note that, with reference to the mathematical materials/notations given in Ref. \cite{Gazeau2022}, the constant polarization five-vector $Z = Z_\alpha$ carries the five-dimensional (nonunitary!) representation $(n_1=0,n_2=1)$ of the dS group. It then follows that the expression ${\cal S} \overline{Z} \widetilde{K}$ is the key piece in reducing the tensor product $(n_1=0,n_2=1)\otimes\Delta(p=2-1,q=1)$.] In view of the commutation relation (\ref{recur1}), we now encounter two new elements $D_2$ and $\theta$, which respectively make a symmetric transverse rank-$2$ tensor field ${\cal K}$ from a rank-$1$ transverse tensor field ${K}$ (as $D_2{K}$) and a rank-$0$ tensor field $\phi$ (as $\theta\phi$). The commutation relations between $Q_2$ and $D_2 {K}$ and of course $\theta\phi$ respectively read:
\begin{eqnarray}
\label{recur2} Q_2D_2{K} &=& D_2 Q_1 {K}\,, \\
\label{recur3} Q_2\theta\phi &=& \theta Q_0\phi\,.
\end{eqnarray}
The three independent elements ${\cal S}\overline{Z} \widetilde{K}$, $D_2{K}$, and $\theta\phi$ thus form a closed family under the action of $Q_2$. Then, the symmetric transverse rank-$2$ tensor field ${\cal K}$ can be expressed in a dS-invariant way in terms of $\widetilde{K}$, ${K}$, and $\phi$ as:
\begin{eqnarray}\label{recuform}
{\cal K} = {\cal S}\overline{Z} \widetilde{K} + D_2{K} + \theta\phi \,.
\end{eqnarray}

In the recurrence formula, the transversality of ${\cal K}$ ($x\cdot{\cal{K}}=0$) is guaranteed by the transversality of the rank-$1$ tensor fields $\widetilde{K}$ and ${K}$ ($x\cdot \widetilde{K} = 0 = x\cdot {K}$). The tensor field (\ref{recuform}) is also supposed to verify the requirements of homogeneity and divergencelessness. Moreover, as a byproduct of these requirements, it needs to be traceless, that is, ${\cal K}^\prime = 2\big( Z\cdot \widetilde{K} + D_1\cdot{K} + 2\phi \big) =0$. On the other hand, the tensor field (\ref{recuform}) has to verify the field equation (\ref{Field Eq. Gen2'}), which implies that its ingredients $\widetilde{K}$, ${K}$, and $\phi$ have to respectively verify:
\begin{eqnarray}
&\label{aaa} Q_1 \widetilde{K} = 0\,,&\\
&\label{bbb} \left( Q_1 +4 \right){K} - 2R^{-2}(Z\cdot x)\widetilde{K} = 0\,,&\\
&\label{ccc} \left( Q_0+4 \right)\phi + 4 (Z\cdot \widetilde{K}) = 0\,.&
\end{eqnarray}
Note that, to get the above equations, we have used the identities (\ref{recur1})-(\ref{recur3}).

The crucial observation to make in this context is that the recurrence formula (\ref{recuform}) must have a group-theoretical interpretation. In this sense, let us delve more deeply into the recurrence formula (\ref{recuform}). Technically, it displays the reduction, through the leading term ${\cal S}\overline{Z}\widetilde{K}$, of the tensor product $(n_1=0,n_2=1)\otimes\Delta(p=2-1,q=1)$ which contains the UIR $\Pi^{\pm}_{2,1}$. Accordingly, $\widetilde{K}$, verifying $Q_1 \widetilde{K}=0$, has to obey the divergenceless condition $\partial\cdot \widetilde{K}=\overline\partial\cdot \widetilde{K}=0$ to be a carrier state for the UIR $\Pi^{\pm}_{1,1}$, characterized by $\Delta(p=1,q=1)$. But, once one imposes such a condition on $\widetilde{K}$, the solution to $Q_1 \widetilde{K}=0$ leads to a singularity of the type $1/\langle Q_1\rangle$, where $\langle Q_1\rangle$, being the quadratic Casimir eigenvalue associated with the UIR $\Pi^{\pm}_{1,1}$, is equal to zero (see Ref. \cite{Massless 1} for the details). To get rid of this singularity, the divergenceless condition on $\widetilde{K}$ ($\partial\cdot \widetilde{K}=0$), must be relaxed. In other words, one must solve the field equation $Q_1 \widetilde{K} =0$ in a larger space which includes one more degree of freedom due to the $\partial\cdot \widetilde{K} \neq 0$ types of solutions. Consequently, the equation $Q_1 \widetilde{K} =0$ turns into a gauge-invariant one $Q_1 \widetilde{K} + D_1 \partial\cdot \widetilde{K} = 0$ in such a way that $\widetilde{K}\mapsto \widetilde{K}+D_1\widetilde{\phi}^{}_g$ is a solution to the field equation for any scalar field $\widetilde{\phi}_g$ as far as $\widetilde{K}$ is. Then, there are three main types of solutions for $\widetilde{K}$, namely, gauge solutions, physical solutions which are divergenceless, and solutions that are not divergenceless. After the gauge-fixing procedure, the explicit form of the solution reads as \cite{Massless 1}:
\begin{eqnarray}\label{K solution}
\widetilde{K} &=& \overline{\widetilde{Z}}\widetilde{\phi} - \frac{\widetilde{\lambda}}{2(1-\widetilde{\lambda})} D_1 \left(R^{-2}(\widetilde{Z}\cdot x)\widetilde{\phi} +\widetilde{Z}\cdot\overline\partial\widetilde{\phi} \right) \nonumber\\
&& + \underline{\underline{\frac{2-3\widetilde{\lambda}}{1-\widetilde{\lambda}} R^{-2}D_1 Q_0^{-1} (\widetilde{Z}\cdot x)\widetilde{\phi}}} + D_1\widetilde{\phi}^{}_g\,,\quad
\end{eqnarray}
where $\widetilde{Z}$ denotes another constant polarization five-vector, $\widetilde{\phi}$ the dS massless conformally coupled scalar field obeying the equation:
\begin{eqnarray}\label{Confomally Eq.}
(Q_0-2)\widetilde{\phi}=0\,,
\end{eqnarray}
and corresponding to the scalar complementary (massless) UIR ${U}^{\mbox{\small{cs}}}_{0, \frac{1}{2}}$, and finally $\widetilde{\lambda}\big(\neq 1\big)$ the gauge-fixing parameter. Note that: (i) After the gauge fixing procedure (for $\widetilde{\lambda}\neq 1$), the gauge field $\widetilde{\phi}^{}_g$ is determined by the equation $Q_0 \widetilde{\phi}^{}_g = 0$, which means that $\widetilde{\phi}^{}_g$ is the minimally coupled scalar field corresponding to the scalar discrete UIR $\Pi_{1,0}$. (ii) The term distinguished above by drawing a double line below is responsible for the appearance of logarithmic divergences in the field solutions which implies reverberation inside the light cone. Accordingly, contrary to the Minkowskian flat case, the ``minimal" (or optimal) choice of $\widetilde{\lambda}$ is not zero. It is clearly $\widetilde{\lambda}=\frac{2}{3}$, which eliminates the logarithmic divergent term. (iii) The solution (\ref{K solution}), instead of the UIR $\Pi^{\pm}_{1,1}$, carries an indecomposable representation of the dS group containing $\Pi^{\pm}_{1,1}$ as its central (physical) part. (iv) The physical part of the solution (\ref{K solution}), carrying the representation $\Pi^{\pm}_{1,1}$, is obtained by imposing the divergenceless condition ($\partial\cdot \widetilde{K}=0$) on (\ref{K solution}) and eliminating the gauge solution $D_1 \widetilde{\phi}_g$:
\begin{eqnarray}\label{partial K}
\partial\cdot \widetilde{K} &=& \partial\cdot\overline{\widetilde{Z}}\widetilde{\phi} + \frac{\widetilde{\lambda}}{2(1-\widetilde{\lambda})} \, Q_0 \left(R^{-2}(\widetilde{Z}\cdot x)\widetilde{\phi} +\widetilde{Z}\cdot\overline\partial\widetilde{\phi} \right) \nonumber\\
&& - \frac{2-3\widetilde{\lambda}}{1-\widetilde{\lambda}} R^{-2} (\widetilde{Z}\cdot x)\widetilde{\phi} \nonumber\\
&=& \frac{1}{1-\widetilde{\lambda}} \left( (2+\tau) R^{-2}(\widetilde{Z}\cdot x) + \tau \frac{\widetilde{Z}\cdot\xi}{x\cdot\xi} \right)\widetilde{\phi}\,,
\end{eqnarray}
where, above, we have invoked the \emph{scalar plane wave} (see Appendix \ref{App. Plane Waves}) solutions to Eq. (\ref{Confomally Eq.}):
\begin{eqnarray}\label{gov}
\widetilde{\phi}(x) = \left(\frac{x\cdot\xi}{R}\right)^\tau , \quad \tau=-1,-2\,,
\end{eqnarray}
where $x$ and $\xi$ respectively live in $M_R$ and in the null-cone $C$ in $\mathbb{R}^5$:
\begin{eqnarray}
C = \Big\{ \xi\in \mathbb{R}^5 \; ; \; (\xi)^2 \equiv \xi\cdot\xi = \eta^{}_{\alpha\beta}\xi^\alpha\xi^\beta =0 \Big\}\,.
\end{eqnarray}
It is evident from Eq. (\ref{partial K}) that the solutions satisfying the divergencelessness condition are not determined by a particular choice of $\widetilde{\lambda}$ (for $\widetilde{\lambda}\neq 1$). Hence, one has the freedom to select the specific value $\widetilde{\lambda}=\frac{2}{3}$. Importantly, the solutions fulfilling the divergencelessness requirement are characterized by the following conditions:
\begin{eqnarray}\label{Phys. conditions}
\tau=-2 \quad \mbox{and} \quad \widetilde{Z}\cdot\xi=0\,.
\end{eqnarray}
Therefore, taking into account the above conditions and substituting the solutions (\ref{gov}) into (\ref{K solution}), the plane-wave reading of the physical part of the solution is obtained as:
\begin{align}\label{K solution phys.}
\widetilde{K} \equiv \widetilde{K}(x,\xi; \widetilde{Z}) =&\; 2 \left( \overline{\widetilde{Z}} - \frac{\widetilde{Z}\cdot x}{x\cdot\xi} \;\overline{\xi} \right) \left(\frac{x\cdot\xi}{R}\right)^{-2}\nonumber\\
\equiv&\; 2\, \varepsilon(x,\xi; \widetilde{Z})\left(\frac{x\cdot\xi}{R}\right)^{-2}, \quad \partial\cdot \widetilde{K} = 0\,.
\end{align}
One notices that $\xi\cdot \varepsilon(x,\xi; \widetilde{Z}) = \overline{\xi}\cdot \varepsilon(x,\xi; \widetilde{Z}) = 0$, as $\xi\cdot\widetilde{Z}=0$. [For more details, readers are referred to Ref. \cite{Massless 1}.]

Systematically (as it will be accurately clarified in the sequel), having the above scenario for the vector field $\widetilde{K}$ entails the traceless (${\cal K}^\prime \neq 0$) and the divergenceless ($\partial\cdot {\cal{K}} \neq 0$) conditions being relaxed from the very beginning in the case of the tensor field ${\cal K}$. Of course, to restrict the relaxed degrees of freedom to $1$, we also impose:
\begin{eqnarray}\label{Constraint}
\partial_2\cdot {\cal{K}} = \frac{1}{2} \overline\partial{\cal K}^\prime\,,
\end{eqnarray}
where `$\partial_2\cdot$' is called the generalized divergence on the dS hyperboloid and technically defined as $\partial_2\cdot {\cal{K}} = \partial\cdot{\cal{K}} - R^{-2}x{\cal{K}}^\prime -\frac{1}{2}\overline\partial{\cal{K}}^\prime$. The field equation (\ref{Field Eq. Gen2'}) then turns into the following gauge-invariant one:
\begin{eqnarray}
\left( Q_2 +4 \right) {\cal{K}} + D_2\partial_2\cdot{\cal{K}} - \theta {\cal{K}}^\prime = 0\,,
\end{eqnarray}
such that:
\begin{eqnarray}\label{Gauge}
{\cal K}\mapsto {\cal K} + D_2D_1\phi^{}_g - 2R^2 \theta \phi^{}_g\,,
\end{eqnarray}
represents a solution to the field equation as far as ${\cal K}$ does ($\phi^{}_g$ being an arbitrary dS scalar field). The latter point can be easily checked by using the identities (\ref{recur2}), (\ref{recur3}), and:
\begin{eqnarray}\label{Identities}
\partial_2\cdot\theta\phi^{}_g &=& -R^{-2}D_1 \phi^{}_g\,,\\
Q_1 D_1 \phi^{}_g &=& D_1 Q_0 \phi^{}_g\,, \\
\partial_2\cdot D_2D_1 \phi^{}_g &=& -(Q_1 +6) D_1 \phi^{}_g \nonumber\\
&=& - D_1 (Q_0 +6) \phi^{}_g\,.
\end{eqnarray}
Now, we introduce a gauge fixing parameter $\lambda$. The field equation then reads:
\begin{eqnarray}\label{Field Eq. Gen2+gauge}
\left( Q_2 +4 \right) {\cal{K}} + \lambda D_2\partial_2\cdot{\cal{K}} - \lambda \theta {\cal{K}}^\prime = 0\,,
\end{eqnarray}
while we have in mind the constraint (\ref{Constraint}).

\subsubsection{Precision on the field equation (\ref{Field Eq. Gen2+gauge})}
Substituting the generic solution (\ref{recuform}) into (\ref{Field Eq. Gen2+gauge}), we get three counterparts of Eqs. (\ref{aaa})-(\ref{ccc}) respectively:
\begin{eqnarray}\label{1}
Q_1 \widetilde{K} + \lambda D_1 \partial \cdot \widetilde{K} = 0\,,
\end{eqnarray}
which is exactly the equation that we have already expected for the transitive vector field $\widetilde{K}$, and:
\begin{align}\label{2}
(Q_1 + 4){K} = 2 R^{-2}(Z\cdot x)\widetilde{K} + \lambda\overline{Z}(\partial\cdot \widetilde{K}) - \frac{1}{2}\lambda \overline\partial{\cal K}^\prime\,,
\end{align}
\begin{eqnarray}\label{3}
(Q_0+4)\phi = - 4 (Z\cdot \widetilde{K}) - 2\lambda (Z\cdot x)\partial \cdot \widetilde{K} + \lambda{\cal K}^\prime \,,
\end{eqnarray}
where $\mathfrak{T} Z\cdot\overline\partial \widetilde{K} \equiv Z\cdot\overline\partial \widetilde{K} - R^{-2} x(Z\cdot \widetilde{K})$. Note that to obtain the above equations, besides the identities given so far, we have also used:
\begin{eqnarray}
\partial_2\cdot {\cal S} \overline{Z}\widetilde{K} &=& \mathfrak{T} Z\cdot\overline\partial \widetilde{K} + \overline{Z}\partial\cdot \widetilde{K} - \overline\partial (Z\cdot \widetilde{K}) \nonumber\\
&&+ 5R^{-2} (Z\cdot x) \widetilde{K} \,,
\end{eqnarray}
and also:
\begin{eqnarray}
&\mathfrak{T} Z\cdot\overline\partial \widetilde{K} + \overline{Z}\partial\cdot \widetilde{K} - \overline\partial (Z\cdot \widetilde{K}) + 5R^{-2}(Z\cdot x)\widetilde{K} &\nonumber\\
& - (Q_1+6){K} - \overline\partial\phi=\frac{1}{2} \overline\partial {\cal K}^\prime \,.&
\end{eqnarray}
The latter equation is actually obtained by substituting the generic solution (\ref{recuform}) into the constraint (\ref{Constraint}) (recall that ${\cal K}^\prime = 2Z\cdot \widetilde{K} + 2D_1\cdot {K} + 4\phi$).

Here, we would like to point out that the above equations, as already expected, are invariant under the transformations ${K}\mapsto{K}+D_1\phi^{}_g$ and $\phi\mapsto\phi-2R^2\phi^{}_g$, issued from the gauge transformation (\ref{Gauge}), such that: if $\lambda= 1$, the scalar field $\phi^{}_g$ remains arbitrary (except for the restrictions imposed by ordinary differentiability prerequisites); if $\lambda\neq 1$, $\phi^{}_g$ is restricted by $(Q_0+4)\phi^{}_g=0$.

\subsection{The Gupta-Bleuler triplet}
\emph{``The appearance of [the Gupta-Bleuler] triplet seems to be universal in gauge theories, and crucial for quantization"} \cite{Binegar}.

The formalism presented above is noteworthy in that it enables us to establish the Gupta-Bleuler triplet for the dS partially massless graviton field, analogous to that for the electromagnetic field in Minkowski spacetime. To do so, first, considering the field equation (\ref{Field Eq. Gen2+gauge}), on one hand and on the other hand, the mathematical materials given in Ref. \cite{GazeauHansMurenzi}, we put forward the corresponding dS-invariant bilinear form (inner product) on the space of solutions:
\begin{eqnarray} \label{inner product gen.}
\langle {\cal K}_1,{\cal K}_2\rangle &=& \mathrm{i} R^2 \int_{\mathbb{S}^3,\rho=0} \Big[({{\cal K}_1})^\ast\cdot\cdot\partial_\rho{{\cal K}_2} \nonumber\\
&& \quad\quad\quad - 2 \lambda \big((\partial_\rho x)\cdot{({{\cal K}_1})}^\ast\big)\cdot(\partial\cdot{{\cal K}_2}) \nonumber\\
&& \quad\quad\quad - (1^\ast \leftrightharpoons 2)\Big] \mathrm{d}\Omega\,,\quad
\end{eqnarray}
where ${\cal{K}}_1$ and ${\cal{K}}_2$ stand for two modes in the space of solutions, and $\mathrm{d}\Omega$ for the invariant measure on $\mathbb{S}^3$. Note that, above, we have employed a system of bounded global intrinsic coordinates ($X^\mu$) well suited for describing a bounded copy of dS spacetime, that is, $\mathbb{S}^3\times\left(-\frac{\pi}{2},\frac{\pi}{2}\right)$. This coordinate system, which is known in the literature under the name of conformal coordinates, is characterized by:
\begin{eqnarray}\label{conformal coordinates}
x = \big(x^0 = R \tan\rho , R (\cos\rho)^{-1} u \big)\,,
\end{eqnarray}
where $-\frac{\pi}{2} < \rho < \frac{\pi}{2}$ and $u \in \mathbb{S}^3$. For the modes that obey the divergenceless and/or traceless condition (recall that the constraint (\ref{Constraint})), the above bilinear form becomes $\lambda$ independent and of Klein-Gordon type:
\begin{align} \label{inner product phys.}
\langle {\cal K}_1,{\cal K}_2\rangle = \mathrm{i} R^2 \int_{\mathbb{S}^3,\rho=0} \Big[({{\cal K}_1})^\ast\cdot\cdot\partial_\rho{{\cal K}_2} - (1^\ast \leftrightharpoons 2) \Big] \mathrm{d}\Omega\,.
\end{align}

We now elaborate the Gupta-Bleuler triplet $V_g \subset V \subset V_\lambda$ which carries the dS indecomposable group representation structure corresponding to our problem:
\begin{itemize}
\item{$V_\lambda$ is the space of all square-integrable (according to (\ref{inner product gen.})) solutions to the field equation (\ref{Field Eq. Gen2+gauge}). In $V_\lambda$, the inner product is indefinite. This means that $V_\lambda$ includes negative norm solutions as well.}

\item{$V$ is the space of the divergenceless and/or traceless solutions (recall that the constraint (\ref{Constraint})). It forms a closed subspace of $V_\lambda$ and, according to the field equation (\ref{Field Eq. Gen2+gauge}), is clearly $\lambda$ independent. Of course, one must notice that the invariant subspace $V$ is not invariantly complemented in $V_\lambda$. The inner product in $V$ is semidefinite.}

\item{$V_g$ is the space of the gauge solutions of the form $ {\cal K}_g = D_2D_1\phi^{}_g - 2R^2 \theta \phi^{}_g $. It forms a closed, \emph{null-norm}\footnote{Every member of the gauge subspace $V_g$ is orthogonal to all members of $V$ including itself!} subspace of $V$. Of course, again, the invariant subspace $V_g$ is not invariantly complemented in $V$.}
\end{itemize}

Below, we delve more deeply into the characteristics of the states belonging to the subspace $V_g$ and the quotient spaces $V/V_g$ and $V_\lambda/V$. We also make explicit the indecomposable group representation structure that is carried by the Gupta-Bleuler triplet $V_g \subset V \subset V_\lambda$.

\subsubsection{Gauge states space; $V_g$}
Considering the gauge solutions of the form $ {\cal K} = {\cal K}_g \equiv D_2D_1\phi^{}_g - 2R^2 \theta \phi^{}_g $, the field equation (\ref{Field Eq. Gen2+gauge}) reduces to:
\begin{eqnarray}
(1-\lambda) (D_2D_1 - 2R^2\theta) (Q_0 +4)\phi^{}_g = 0\,.
\end{eqnarray}
It follows that:
\begin{itemize}
\item{If $\lambda=1$, the scalar field $\phi^{}_g$ remains arbitrary, with of course mild differentiability prerequisites. The gauge states space $V_g$ is then characterize by the symmetric transverse rank-2 tensors of the form $ {\cal K}_g = D_2D_1\phi^{}_g - 2R^2 \theta \phi^{}_g $, with an arbitrary differentiable scalar field $\phi^{}_g$.}

\item{If $\lambda\neq 1$, the scalar field $\phi^{}_g$ is restricted by the equation $(Q_0 +4)\phi^{}_g = 0$ (possibly up to the addition of a specific solution to the inhomogeneous equation $(Q_0 +4)\phi^{}_g = \psi^{}_g$, such that $(D_2D_1 - 2R^2 \theta)\psi^{}_g=0$). Moreover, since:
    \begin{eqnarray}
    L^{(2)}_{\alpha\beta} {\cal K}_g &=& L^{(2)}_{\alpha\beta} \Big( D_2D_1 \phi^{}_g - 2R^2 \theta \phi^{}_g\Big) \nonumber\\
    &=& D_2L^{(1)}_{\alpha\beta}D_1\phi^{}_g - 2R^2 \theta M^{}_{\alpha\beta} \phi^{}_g \nonumber\\
    &=& D_2D_1 M^{}_{\alpha\beta}\phi^{}_g - 2R^2 \theta M^{}_{\alpha\beta}\phi^{}_g \nonumber\\
    &=& \Big( D_2D_1 - 2R^2 \theta \Big) M^{}_{\alpha\beta}\phi^{}_g\,,
    \end{eqnarray}
    the gauge solutions ${\cal K}_g$ do not carry any spin; ${\cal K}_g$s are entirely characterized by their scalar content $\phi^{}_g$ corresponding to the scalar discrete UIR $\Pi_{p=2,q=0}$ (since, we have $(Q_0 +4)\phi^{}_g = 0$). It is also useful to point out that the trace of the gauge solutions reads as:
    \begin{eqnarray}
    {\cal K}^\prime_g = -2R^2 (Q_0 + 4) \phi^{}_g\,.
    \end{eqnarray}
    Therefore, as far as $\lambda\neq 1$, and consequently, $\phi^{}_g$ is restricted by $(Q_0 +4)\phi^{}_g = 0$, the gauge solutions are tracelessness. In this context, one can simply show that the gauge solutions are also divergencelessness: 
    \begin{align}
    \partial_2\cdot{\cal{K}}_g = -D_1(Q_0+4)\phi^{}_g = 0\,.
    \end{align}}
\end{itemize}

\subsubsection{Physical states space (central part); $V/V_g$}
The physical state space is $\lambda$ independent. It consists of the solutions ${\cal K} = {\cal{K}}_{phys}$, which are defined up to the gauge solutions ${\cal K}_g$ and obtained by imposing the divergenceless and/or traceless condition on the field equation (\ref{Field Eq. Gen2+gauge}):
\begin{eqnarray}\label{Field Eq. Phys}
\left( Q_2 +4 \right) {\cal{K}}_{phys} = 0\,.
\end{eqnarray}
The dS group acts on physical states space $V/V_g$ by the UIR $\Pi^\pm_{2,1}$. In the sequel, we will show that these modes propagate on the light cone.

\subsubsection{Scalar states space (pure-trace part); $V_\lambda/V$}
First, we merge the field equation (\ref{Field Eq. Gen2+gauge}) with the constraint (\ref{Constraint}), which yields:
\begin{eqnarray}
\left( Q_2 +4 \right) {\cal{K}} + \frac{1}{2} \lambda D_2\overline\partial{\cal K}^\prime - \lambda \theta {\cal{K}}^\prime = 0\,.
\end{eqnarray}
Then, the scalar states (or the pure-trace part)\footnote{Recall that, due to the constraint (\ref{Constraint}), the relaxed degrees of freedom reduce to $1$. This is the fact that assures us that the exterior part of the Gupta-Bleuler triplet (i.e., $V_\lambda/V$) is merely characterized by the pure-trace solutions.} can be achieved in two steps: first, by setting $ {\cal{K}} = {\cal K}_{pt} \equiv \frac{1}{4} \theta {\cal K}^\prime$ in the yielded equation:
\begin{eqnarray}
\frac{1}{4}\theta (Q_0 +4){\cal K}^\prime + \frac{1}{2} \lambda D_2\overline\partial{\cal K}^\prime - \lambda \theta{\cal K}^\prime = 0\,,
\end{eqnarray}
and subsequently, by taking the trace of it:
\begin{eqnarray}\label{trace constraint}
(1-\lambda)(Q_0 +4){\cal K}^\prime = 0 \,.
\end{eqnarray}
Here, again, two distinguished cases appear:
\begin{itemize}
\item{If $\lambda=1$, ${\cal K}^\prime$ remains unrestricted (except for the restrictions imposed by ordinary differentiability prerequisites). Then, one naturally loses the ability to restrain the space $V_\lambda$.}

\item{If $\lambda\neq 1$, ${\cal K}^\prime$ is restricted by $(Q_0 +4){\cal K}^\prime = 0$, which implies that ${\cal K}^\prime$ also corresponds to the scalar discrete UIR $\Pi_{p=2,q=0}$.}
\end{itemize}

\subsubsection{Indecomposable group representation structure}
The indecomposable group representation structure, carried by the above Gupta-Bleuler triplet, associated with the dS partially massless graviton field, then reads:
\begin{eqnarray}\label{indecomposable}
\underbrace{{\Pi_{2,0}}}_{V_\lambda/V}\;\;\dashrightarrow\;\; \underbrace{\Pi_{2,1}^{\pm}}_{V/V_g}\;\;\dashrightarrow\;\; \underbrace{{\Pi_{2,0}}}_{V_g}\,,
\end{eqnarray}
where the arrows stand for the leaks under the dS group action. Note that the UIRs $\Pi^\pm_{2,1}$ and $\Pi_{2,0}$ share the same Casimir eigenvalue, and hence, they are Weyl equivalent.\footnote{In general, a necessary condition for combining different representations to construct an indecomposable group representation structure is that they are Weyl equivalent.}

\subsection{Field equation solutions and de Sitter waves}

\subsubsection{Field equation solutions}
We now solve the field equation (\ref{Field Eq. Gen2+gauge}), for $\lambda\neq 1$, with the constraint (\ref{Constraint}). For the sake of simplicity, we merely restrict our attention to the physical sector of the vector field $\widetilde{K}$  (see Eq. (\ref{K solution phys.})), which is free of singularities, and realized by imposing the divergenceless condition ($\partial\cdot \widetilde{K}=0$) on (\ref{K solution}) and eliminating the respective gauge solution $D_1 \widetilde{\phi}_g$. Then, we get the simplest forms of Eqs. (\ref{1})-(\ref{3}), respectively, as:
\begin{eqnarray}
\label{11} Q_1 \widetilde{K} &=& 0\,,\\
\label{22} (Q_1 + 4){K} &=& 2 R^{-2}(Z\cdot x)\widetilde{K} - \frac{1}{2}\lambda \overline\partial{\cal K}^\prime\,,\\
\label{33} (Q_0+4)\phi &=& - 4 (Z\cdot \widetilde{K}) + \lambda{\cal K}^\prime \,.
\end{eqnarray}

\emph{\textbf{Important remark:}} Certainly, it should be noted that the imposition of the condition $\partial\cdot \widetilde{K}=0$ does not change the underlying physics of the problem, in the sense that: (i) When comparing the aforementioned set of equations with its previous form, the imposition of $\partial\cdot \widetilde{K}=0$ merely confines the nonphysical (dependent on $\lambda$) components of the equations, consequently merely affecting the nonphysical (dependent on $\lambda$) part of the corresponding solution. (ii) This simplified set of equations, similar to its previous version, remains unchanged under the transformations ${K}\mapsto{K}+D_1\phi^{}_g$ and $\phi\mapsto\phi-2R^2\phi^{}_g$, issued from the gauge transformation (\ref{Gauge}), such that again: if $\lambda= 1$, the scalar field $\phi^{}_g$ remains arbitrary (except for the restrictions imposed by ordinary differentiability prerequisites); if $\lambda\neq 1$, $\phi^{}_g$ is restricted by $(Q_0+4)\phi^{}_g=0$. Therefore, the imposition of $\partial\cdot \widetilde{K}=0$ does not change the gauge-invariant property of the theory, as well.

We begin with Eq. (\ref{33}), based upon which we have:
\begin{eqnarray}\label{33'}
\phi &=& (Q_0+4)^{-1} \Big(- 4 (Z\cdot \widetilde{K}) + \lambda{\cal K}^\prime\Big) \nonumber\\
&=& - \frac{2}{3} (Z\cdot \widetilde{K}) + \lambda (Q_0+4)^{-1} {\cal K}^\prime\,.
\end{eqnarray}
This result can be verified by utilizing the identity:
\begin{eqnarray}
\left( Q_0+4 \right)(Z\cdot \widetilde{K}) = 6(Z\cdot \widetilde{K})\,,
\end{eqnarray}
which is obtained by using Eq. (\ref{11}) along with the identity:
\begin{align}\label{Q_1}
Q_1 \widetilde{K} = (Q_0-2)\widetilde{K} + 2x\partial\cdot \widetilde{K} - 2\partial x\cdot \widetilde{K}\,.
\end{align}
while we have in mind that $x\cdot \widetilde{K}=0=\partial\cdot \widetilde{K}$. Note that the last term in (\ref{33'}) bears a logarithmic singularity, since $\left( Q_0+4 \right){\cal K}^\prime=0$ (see Eq. (\ref{trace constraint}) and its subsequent arguments).

Now, we deal with Eq. (\ref{22}). Let $\mathfrak{S}$ be a three-dimensional space spanned by linear combinations of the following set of three basic functions:
\begin{eqnarray}\label{S}
\mathfrak{S} \ni \big[s^{}_1,s^{}_2,s^{}_3\big] &=& s^{}_1 R^{-2}(Z\cdot x)\widetilde{K} + s^{}_2 \mathfrak{T}(Z\cdot\overline\partial)\widetilde{K} \nonumber\\
&& + s^{}_3 \overline\partial (Z\cdot \widetilde{K}) \,.
\end{eqnarray}
This space is invariant under the action of $(Q_1 + 4)$:
\begin{eqnarray}
(Q_1 + 4) R^{-2}(Z\cdot x)\widetilde{K} &=& \big[0,-2,0\big]\,,\\
(Q_1 + 4) \mathfrak{T}(Z\cdot\overline\partial)\widetilde{K} &=& \big[4,6,2\big]\,,\\
(Q_1 + 4) \overline\partial (Z\cdot \widetilde{K}) &=& \big[0,0,6\big]\,.
\end{eqnarray}
Note that to get the above equations, besides Eqs. (\ref{11}) and (\ref{Q_1}), we have also used the following identities:
\begin{eqnarray}
x (Q_0 - 2) &=& (Q_0+2)x + 2R^2 \overline{\partial}\,,\\
\overline{\partial} (Q_0 - 2) &=& (Q_0-8) \overline{\partial} - 2R^{-2} (Q_0+4)x\,.
\end{eqnarray}

On the other hand, $\overline\partial {\cal{K}}^\prime$ itself forms an invariant one-dimensional space under the action of $(Q_1 + 4)$ such that $(Q_1 + 4)\overline\partial {\cal K}^\prime = 0$ (which is a byproduct of Eq. (\ref{trace constraint}) and its subsequent arguments).

With these three and one-dimensional spaces in mind, which remain invariant under the $(Q_1 + 4)$ action, we rewrite Eq. (\ref{22}) as:
\begin{align}\label{22'}
(Q_1 + 4){K} = \big[2,0,0\big] - \frac{1}{2}\lambda \overline\partial{\cal K}^\prime\,.
\end{align}
The general solution to (\ref{22'}) can be divided into two sectors:
\begin{align}\label{22''}
{K} = \big[\mathrm{s}^{}_1,\mathrm{s}^{}_2,\mathrm{s}^{}_3\big] - \frac{1}{2}\lambda (Q_1 + 4)^{-1} \overline\partial{\cal K}^\prime\,,
\end{align}
such that the first sector is obtained by the following system:
\begin{eqnarray}\label{22'''}
\left(\begin{array}{cccccccc}
0 & 4 & 0 \\
-2 & 6 & 0 \\
0 & 2 & 6 \\
\end{array}\right)
\left(\begin{array}{cccccccc}
\mathrm{s}^{}_1 \\
\mathrm{s}^{}_2 \\
\mathrm{s}^{}_3 \\
\end{array}\right) =
\left(\begin{array}{cccccccc}
2 \\
0 \\
0 \\
\end{array}\right)\,,
\end{eqnarray}
which yields:
\begin{eqnarray}
{K} = \left[\frac{3}{2},\frac{1}{2},-\frac{1}{6}\right] - \frac{1}{2}\lambda (Q_1 + 4)^{-1} \overline\partial{\cal K}^\prime \,,
\end{eqnarray}
where the last term bears a logarithmic singularity.

Accordingly, a general solution to the field equation (\ref{Field Eq. Gen2+gauge}), with the constraint (\ref{Constraint}), can be expressed as follows, after making straightforward adjustments:
\begin{align}\label{Solu gen.}
&{\cal K} = - \frac{2}{3} \theta (Z\cdot \widetilde{K}) + {\cal S}\overline Z \widetilde{K} + \frac{3}{2} D_2 R^{-2}(Z\cdot x)\widetilde{K} \nonumber\\
& \quad + \frac{1}{2} D_2 \mathfrak{T}(Z\cdot\overline\partial)\widetilde{K} -\frac{1}{6} D_2 \overline\partial (Z\cdot \widetilde{K}) \nonumber\\
& \quad - \frac{\lambda}{2} R^{-2} \Big( D_2 (Q_1 + 4)^{-1} D_1{\cal K}^\prime - 2R^2 \theta (Q_0+4)^{-1} {\cal K}^\prime \Big)\,. \quad\quad
\end{align}
It should be noted that the gauge solution ${\cal K}_g = D_2D_1\phi^{}_g - 2R^2 \theta \phi^{}_g$ is coupled to the $\lambda$-dependent part of the solution (given in the last line of Eq. (\ref{Solu gen.})). Furthermore, given the aforementioned solution, it is evident that the optimal choice for the gauge-fixing parameter is $\lambda=0$, which effectively removes all singular terms. Consequently, from now on, we will proceed with this optimal selection of the gauge-fixing parameter.

At this point, we are required to present the interpretation of the solution (\ref{Solu gen.}) in terms of a plane wave. To aid our reasoning, let us begin by revisiting the plane-wave explanation of the transitive and divergenceless vector field $\widetilde{K}$, as outlined in Eq. (\ref{K solution phys.}). Following that, it is worth recalling:
\begin{align}\label{K solution new}
\widetilde{K}(x) =& 2\, \varepsilon(x,\xi;{\widetilde{Z}}^{\kappa}_{})\left(\frac{x\cdot\xi}{R}\right)^{-2} \nonumber\\
=& 2 \left( {\overline{\widetilde{Z}}}^{\kappa} - \frac{{\widetilde{Z}}^{\kappa}_{} \cdot x}{x\cdot\xi} \;\overline{\xi} \right) \left(\frac{x\cdot\xi}{R}\right)^{-2}.
\end{align}
Note that when choosing constant five-vectors, such as $\widetilde{Z}_\alpha$, they are commonly labeled with $\kappa=0,1,2,3$.\footnote{This actually arises from the fact that despite being expressed as five-component vectors, objects in dS spacetime possess only four independent components.} These vectors then can be denoted as ${\widetilde{Z}}^{(\kappa)}_{\alpha}$ or, for the sake of convenience in expressions, as ${\widetilde{Z}}^{\kappa}_{\alpha}$. Now, let us impose the following conditions on ${\widetilde{Z}}^{\kappa}_{\alpha}$:
\begin{eqnarray}
\label{conditions on Z1} {\widetilde{Z}}^{\kappa}_{} \cdot {\widetilde{Z}}^{\kappa^\prime}_{} &=& \eta_{}^{\kappa\kappa^\prime}\,, \\
\label{conditions on Z2} \sum_{\kappa=1}^3 {\widetilde{Z}}^{\kappa}_{\alpha} {\widetilde{Z}}^{\kappa}_{\beta} &=& - \eta^{}_{\alpha\beta}\,,\\
\label{conditions on Z3} \sum_{\kappa=1}^3 {\widetilde{Z}}^{\kappa}_{4} {\widetilde{Z}}^{\kappa}_{\mu} &=& 0\,,
\end{eqnarray}
for all $\kappa,\kappa^\prime = 0,1,2,3$. Subsequently, it is evident that the characteristics of the dS polarization vector $\varepsilon(x,\xi;{\widetilde{Z}}^{\kappa}_{}) \equiv \varepsilon^\kappa_{}(x,\xi)$, bear a striking resemblance to the Minkowskian scenario (for the latter, see Ref. \cite{Itzykson}):\footnote{\label{foot xi.xi}Note that although $\xi\cdot\xi=0$, the products ${\xi}\cdot\overline{\xi}$ and $\overline{\xi}\cdot\overline{\xi}$ are not equal to zero. Specifically, we have ${\xi}\cdot\overline{\xi} = \overline{\xi}\cdot\overline{\xi}=\left( \frac{x\cdot\xi}{R} \right)^2$. Similarly, while $\widetilde{Z}\cdot\xi=0$, the dot products ${\widetilde{Z}}\cdot\overline{\xi}$, $\overline{\widetilde{Z}}\cdot{\xi}$, and $\overline{\widetilde{Z}}\cdot\overline{\xi}$ are not zero. In particular, they can be expressed as ${\widetilde{Z}}\cdot\overline{\xi} = \overline{\widetilde{Z}}\cdot{\xi} = \overline{\widetilde{Z}}\cdot\overline{\xi} = \frac{1}{R^2}({\widetilde{Z}}\cdot x)(x \cdot \xi)$.}
\begin{eqnarray}
\label{Hasan}\sum_{\kappa=1}^3\, \varepsilon^\kappa_{\alpha}(x,\xi) \, \varepsilon^{\kappa}_{\beta}(x,\xi) &=& - \left( \theta_{\alpha\beta} - \frac{{\overline{\xi}}_\alpha {\overline{\xi}}_\beta}{(x\cdot\xi/R)^2} \right)\,,\quad\\
\label{Hosein}\varepsilon^\kappa_{}(x,\xi) \cdot \varepsilon^{\kappa^\prime}_{}(x,\xi) &=& {\widetilde{Z}}_{}^\kappa \cdot \varepsilon^{\kappa^\prime}_{}(x,\xi) \nonumber\\
&=& \varepsilon^\kappa_{}(x,\xi) \cdot \varepsilon^{\kappa^\prime}_{}(x^\prime,\xi) \nonumber\\
&=& \widetilde{Z}^{\kappa}_{}\cdot\widetilde{Z}^{\kappa^\prime}_{} \nonumber\\
&=& \eta_{}^{\kappa\kappa^\prime}\,.
\end{eqnarray}

It follows from Eqs. (\ref{Solu gen.}) and (\ref{K solution new}) that the solution to the field equation (\ref{Field Eq. Gen2+gauge}), with the constraint (\ref{Constraint}), takes the form:
\begin{eqnarray}\label{javad}
{\cal K}(x) = {\cal D}(x,\partial;Z,\widetilde{Z}) \left(\frac{x\cdot\xi}{R}\right)^{-2},
\end{eqnarray}
where the operator ${\cal D}(x,\xi;Z,\widetilde{Z})$, for $\lambda=0$, is given by:
\begin{align}\label{D operator}
&{\cal D}(x,\partial;Z,\widetilde{Z}) = 2 \bigg(- \frac{2}{3} \theta (Z\cdot ) + {\cal S}\overline Z + \frac{3}{2} D_2 R^{-2}(Z\cdot x) \nonumber\\
&\hspace{2cm} + \frac{1}{2} D_2 \mathfrak{T}(Z\cdot\overline\partial) -\frac{1}{6} D_2 \overline\partial (Z\cdot) \bigg)\varepsilon(x,\xi;{\widetilde{Z}})\,,
\end{align}
where, for simplicity, we have again excluded the superscript '$\kappa$'. Subsequently, expressing this solution in the following form becomes a matter of straightforward calculations:
\begin{eqnarray}\label{General plane wave}
{\cal K}_{\alpha\beta}(x) = a^{}_0\, {\cal E}_{\alpha\beta}(x,\xi;Z,\widetilde{Z}) \left(\frac{x\cdot\xi}{R}\right)^{-2},
\end{eqnarray}
with $a^{}_0 = 6\,c^{}_0$, where $c^{}_0$ is a normalization constant, and ${\cal E}_{\alpha\beta}(x,\xi;Z,\widetilde{Z})$ represents the corresponding polarization tensor. Here, we consider an explicit realization of the latter, which is attained through the following procedure.

By imposing the conditions stated in (\ref{conditions on Z1})-(\ref{conditions on Z3}), we have managed to partially remove the freedom arising from the introduction of constant vector $\widetilde{Z}$ in our solution. We now proceed with a similar procedure to determine the value of $Z$. Accordingly, if we select $Z$ to be equal to $\widetilde{Z}$ and refer to both as $Z$ hereafter, the polarization tensor ${\cal E}_{\alpha\beta}(x,\xi;Z,\widetilde{Z})\equiv{\cal E}_{\alpha\beta}(x,\xi;Z)$ adopts a straightforward expression such that, using Eq. (\ref{D operator}) as a starting point, we get:
\begin{align}\label{y 1}
&{\cal E}^{}_{\alpha\beta}(x,\xi;Z) \equiv {\cal E}^{\kappa\kappa^\prime}_{\alpha\beta}(x,\xi) \nonumber\\
&\quad\ = \frac{1}{2} \Bigg({\cal S}\, \varepsilon^\kappa_{\alpha}(x,\xi) \,\varepsilon^{\kappa^\prime}_{\beta}(x,\xi) \nonumber\\
&\quad\quad\quad\quad\quad - \frac{2}{3}\left(\theta_{\alpha\beta} - \frac{\overline{\xi}_\alpha \overline{\xi}_\beta}{(x\cdot\xi/R)^2}\right)\varepsilon^\kappa_{}(x,\xi) \cdot\varepsilon^{\kappa^\prime}_{}(x,\xi)\Bigg)\,,
\end{align}
where, again, $\varepsilon(x,\xi;{\widetilde{Z}}^{\kappa}_{}) \equiv \varepsilon^\kappa_{}(x,\xi)$. Considering the identities given (\ref{Hasan}) and (\ref{Hosein}), one can also rewrite the above result as follows:
\begin{eqnarray}\label{y 2}
{\cal E}^{\kappa\kappa^\prime}_{\alpha\beta}(x,\xi) &=& \frac{1}{2} \Bigg({\cal S}\, \varepsilon^\kappa_{\alpha}(x,\xi) \,\varepsilon^{\kappa^\prime}_{\beta}(x,\xi) \nonumber\\
&& \quad\quad + \frac{2}{3} \eta_{}^{\kappa\kappa^\prime} \sum_{\rho=1}^3 \varepsilon^\rho_{\alpha}(x,\xi) \,\varepsilon^{\rho}_{\beta}(x,\xi)\Bigg)\,. \quad\quad
\end{eqnarray}

Here are a few aspects to consider regarding the provided dS tensor wave ${\cal K}_{\alpha\beta}$.

\emph{First:} Evidently, from Eq. (\ref{y 1}), or equivalently (\ref{y 2}), we obtain $\eta_{}^{\alpha\beta}{\cal E}^{}_{\alpha\beta}(x,\xi;Z) \equiv {\cal E}^{\prime}_{}(x,\xi;Z) = 0$. This means that when the gauge-fixing parameter $\lambda$ is set to zero, we are effectively constrained to the traceless and/or divergenceless part of the solution living in the subspace $V$. On the other hand, exploring the ``dot product" of the provided polarization tensor with $\xi$ or $\overline{\xi}$ uncovers another intriguing characteristic of the obtained polarization:
\begin{align}\label{khar}
\xi\cdot{\cal E} = \overline{\xi}\cdot{\cal E} = 0\,.
\end{align}
The crucial point to note here is that the transversality conditions given in (\ref{khar}) are valid only for the physical part of the solution residing in the quotient space $V/V_g$, rather than the entire solution belonging to $V$, which includes both the physical and the gauge parts. [To grasp the concept, one can simply substitute the plane-wave representation of $\phi^{}_g = \left( x\cdot\xi/R\right)^{-4}$, taking into account that $(Q_0+4)\left( x\cdot\xi/R\right)^{-4} = 0$, into the gauge solution ${\cal K}_g = D_2D_1\phi^{}_g - 2R^2 \theta \phi^{}_g$. Additionally, one needs to consider the points mentioned in footnote \ref{foot xi.xi}.] It is also easy to check that the tensor polarization verifies the following identity:
\begin{align}
{\cal E}^{\kappa\kappa^\prime}_{}(x,\xi) \cdot\cdot\, {\cal E}^{\kappa^{\prime\prime}\kappa^{\prime\prime\prime}}_{}(x,\xi) &= {\cal E}^{\kappa\kappa^\prime}_{}(x^\prime,\xi) \cdot\cdot\, {\cal E}^{\kappa^{\prime\prime}\kappa^{\prime\prime\prime}}_{}(x,\xi) \nonumber\\
&= \eta^{\kappa\kappa^{\prime\prime}}_{} \eta^{\kappa^\prime\kappa^{\prime\prime\prime}} + \eta^{\kappa\kappa^{\prime}}_{} \eta^{\kappa^{\prime\prime}\kappa^{\prime\prime\prime}}\,.\quad
\end{align}

\emph{Second:} The dS tensor wave ${\cal K}_{\alpha\beta}(x)$ exhibits a homogeneity property with the degree of $\tau=-2$ when considered on both the null-cone $C$ and the dS submanifold $M_R$ $\big( \subset \mathbb{R}^5\big)$, which is defined by the condition $x\cdot x = -R^2$, where $R$ remains constant. This arises from the homogeneity property of the scalar plane wave $\left(x\cdot\xi/R\right)^{-2}$ (see Appendix \ref{App. Plane Waves}) and the fact that:
\begin{align}
\varepsilon^{\kappa}(x,a\xi) = \varepsilon^{\kappa}(x,\xi)\,, \quad\mbox{and}\quad \varepsilon^{\kappa}(ax,\xi) = \varepsilon^{\kappa}(x,\xi)\,,
\end{align}
which is evident from the definition of $\varepsilon^{\kappa}(x,\xi)$:
\begin{align}
\varepsilon^{\kappa}(x,\xi) = \left( {\overline{Z}}^{\kappa} - \frac{Z^{\kappa}_{} \cdot x}{x\cdot\xi} \;\overline{\xi} \right) = \left( {Z}^{\kappa} - \frac{{Z}^{\kappa}_{} \cdot x}{x\cdot\xi} \;{\xi} \right)\,.
\end{align}

\emph{Third:} Within this framework, it becomes feasible to develop the covariant quantization of the massless partially graviton field by virtue of the closure of ${\cal K}_{\alpha\beta}(x)$ under the dS group action:
\begin{align}\label{dS covariance K}
\Big( U(g) {\cal K} \Big)_{\alpha\beta}(x) =& \,g^\gamma_\alpha g^\delta_\beta {\cal K}_{\gamma\delta}(g^{-1}x) \nonumber\\
=&\, g^\gamma_\alpha g^\delta_\beta\, a^{}_0\, {\cal E}_{\gamma\delta}(g^{-1}x,\xi;Z) \left(\frac{g^{-1}x\cdot\xi}{R}\right)^{-2}\nonumber\\
=&\, a^{}_0\, {\cal E}_{\alpha\beta}(x,g\xi;gZ) \left(\frac{x\cdot g\xi}{R}\right)^{-2}.
\end{align}
This can be easily demonstrated as the vector polarization meets the following condition:
\begin{align}
\varepsilon_\alpha(g^{-1}x,\xi;Z) =& \left( {Z}_\alpha - \frac{Z \cdot g^{-1}x}{g^{-1}x\cdot\xi} \; \xi_\alpha \right) \nonumber\\
=& \left( {Z}_\alpha - \frac{gZ \cdot x}{x\cdot g\xi} \;{\xi}_\alpha \right) \nonumber\\
=&\; (g^{-1})^\gamma_\alpha\, \varepsilon_\gamma(x,g\xi;gZ)\,.
\end{align}

\emph{Fourth:} The final point to be discussed here is that the wave ${\cal K}_{\alpha\beta}(x)$, as a \emph{function} on $M_R$, is only locally defined on connected open subsets of $M_R$ (refer to Appendix \ref{App. Plane Waves}). To ensure a global definition of this solution, as elucidated in Appendix \ref{App. Plane Waves}, an analytical continuation becomes crucial. Consequently, we define the dS tensor wave ${\cal K}_{\alpha\beta}(x,\xi)$ as the boundary value of the analytic continuation of the solution (\ref{General plane wave}) to the forward-tube ${\cal{T}}^+$; for $z\in{\cal{T}}^+ = \left\{ \mathbb{R}^5 + \mathrm{i} \interior{V}^+ \right\} \cap {M}_R^{(\mathbb{C})}$ and $\xi\in C^+$, the following holds:
\begin{eqnarray}\label{z solution}
{K}_{\alpha\beta}(z,\xi) = a^{}_0\, {\cal E}^{\kappa\kappa^\prime}_{\alpha\beta}(z,\xi)\, \left(\frac{z\cdot\xi}{R}\right)^{-2}.
\end{eqnarray}
Then, the corresponding boundary value \emph{(in the distribution sense)} of the above-complexified waves yields a single-valued global plane-wave reading of the solution (\ref{General plane wave}) as:
\begin{align}\label{masud}
&\text{bv}\; {K}_{\alpha\beta}(z,\xi) \equiv {\cal K}_{\alpha\beta}(x,\xi) \nonumber\\
&\quad = a^{}_0\, {\cal E}^{\kappa\kappa^\prime}_{\alpha\beta}(x,\xi)\, \left(\frac{(x+\mathrm{i}y)\cdot\xi}{R}\right)^{-2}\bigg|_{\xi \in C^+,\; y \in \interior{{V}}^+, \; y \rightarrow 0}\,.
\end{align}
For the explicit form of the latter term see Appendix \ref{App. Plane Waves}.

\section{Two-point function and quantum field}\label{Sec. IV}
In this section, we continue with the quantization of the partially massless graviton field in dS spacetime, building upon the concrete quantization procedure that was previously presented and discussed for scalar fields in the earlier works by Bros et al. \cite{GazeauPRL, Bros 2point func}, and subsequently extended by other authors \cite{Massive/Massless 1/2, Massless 1, Massive 1, BehrooziTakook, Massive 2, Massive 3/2, Massless 2, Massless 2', Massless 2'', Bamba 1, dS gravity 1, dS gravity 2, Gupta 2000, deBievre'} to higher spin fields.

\subsection{Two-point function}
In this study, our focus lies on the free field part of the theory, where all truncated correlation functions vanish. Consequently, the complete characterization of the respective QFT is entirely captured by the corresponding Wightman two-point function ${\cal{W}}_{\alpha\beta\alpha^\prime\beta^\prime}(x,x^\prime)$, where $\alpha,\beta,\alpha^\prime,\beta^\prime = 0,1,\,...\,,4$. The two-point function must meet the following criteria:
\begin{itemize}
  \item{\emph{\textbf{Indefinite sesquilinear form:}}
    \begin{align}
    &\int_{M_R\times M_R}f^{\ast\alpha\beta}(x){\cal{W}}_{\alpha\beta\alpha^\prime\beta^\prime}(x,x^\prime)f^{\alpha^\prime\beta^\prime}(x^\prime) \nonumber\\
    &\hspace{3cm} \times\; \mathrm{d}\mu (x) \mathrm{d}\mu (x^\prime)\,,
    \end{align}
    where $f_{\alpha\beta}$ stands for a test function in the space of functions $C^{\infty}$ with compact support in $M_R$, and $\mathrm{d}\mu (x)$ for the invariant measure on $M_R$.}
  \item{\emph{\textbf{Covariance:}}
    \begin{align}
    &(g^{-1})_\alpha^\gamma(g^{-1})_\beta^\delta{\cal{W}}_{\gamma \delta \gamma\prime \delta\prime}(gx,gx^\prime)g_{\alpha^\prime}^{\gamma\prime}g_{\beta^\prime}^{\delta\prime} \nonumber\\
    &\hspace{3cm}= {\cal{W}}_{\alpha\beta\alpha^\prime\beta^\prime}(x,x^\prime)\,,
    \end{align}
    for all $g\in SO_0(1,4)$.}
  \item{\emph{\textbf{Locality:}}
    \begin{eqnarray}\label{locality re.}
    {\cal{W}}_{\alpha\beta\alpha^\prime\beta^\prime}(x,x^\prime)={\cal{W}}_{\alpha^\prime\beta^\prime\alpha\beta}(x^\prime,x)\,,
    \end{eqnarray}
    if $x$ and $x^\prime$ are spacelike separated ($x\cdot x^\prime>-R^{2}$).}
  \item{\emph{\textbf{Index symmetrizer:}}
    \begin{eqnarray}
    {\cal{W}}_{\alpha\beta\alpha^{\prime}\beta^{\prime}}(x,x^{\prime}) &=& {\cal{W}}_{\alpha\beta\beta^{\prime}\alpha^{\prime}}(x,x^{\prime}) \nonumber\\
    &=& {\cal{W}}_{\beta\alpha\alpha^{\prime}\beta^{\prime}}(x,x^{\prime})\,.
    \end{eqnarray}}
  \item{\emph{\textbf{Transversality:}}
    \begin{align}
    x^\alpha{\cal{W}}_{\alpha\beta\alpha^{\prime}\beta^{\prime}}(x,x^{\prime}) = 0 = {x^{\prime\alpha^{\prime}}}{\cal{W}}_{\alpha\beta\alpha^{\prime}\beta^{\prime}}(x,x^{\prime})\,.
    \end{align}}
  \item{\emph{\textbf{Normal analyticity:}} The Wightman two-point function ${\cal{W}}_{\alpha\beta\alpha^\prime\beta^\prime}(x,x^\prime)$ is the boundary value (in the sense of distribution) of a function $W_{\alpha\beta\alpha^\prime\beta^\prime}(z,z^\prime)$ that is analytic in the tuboid domain \cite{GazeauPRL, Bros 2point func}:
      \begin{eqnarray}
      {\cal{T}}^{+(2)}_{} = \Big\{ (z,z^\prime) \; ; \; z\in {\cal{T}}_{}^-, \; z^\prime \in {\cal{T}}_{}^+ \Big\}\,,
      \end{eqnarray}
      where, again, ${\cal{T}}_{}^\pm$ are respectively the forward and backward tubes of ${M}_R^{(\mathbb{C})}$, respectively, defined by:
      \begin{eqnarray}
      {\cal{T}}^\pm = \Big\{ \mathbb{R}^5 + \mathrm{i} \interior{V}^\pm \Big\} \cap {M}_R^{(\mathbb{C})}\,,
      \end{eqnarray}
      and the domains ${\interior{V}}^\pm \equiv \big\{y \in\mathbb{R}^5 \;;\; (y)^2 > 0, y^0 \gtrless 0 \big\}$ originate from the causal structure of ${M}_R$.}
\end{itemize}

Based on the previously mentioned condition of normal analyticity, we can derive the following conclusions \cite{GazeauPRL, Bros 2point func}: (i) The function $W_{\alpha\beta\alpha^\prime\beta^\prime}(z,z^\prime)$ exhibits maximal analyticity, allowing for its analytic continuation to the ``cut domain":
\begin{eqnarray}
\Delta = \Big\{(z,z^\prime)\in {M}_R^{(\mathbb{C})}\times {M}_R^{(\mathbb{C})} \;;\; (z-z^\prime)^2 < 0 \Big\}\,.
\end{eqnarray}
(ii) The ``permuted Wightman two-point function" ${\cal{W}}_{\alpha^\prime\beta^\prime\alpha\beta}(x^\prime,x)$ corresponds to the boundary value of $W_{\alpha\beta\alpha^\prime\beta^\prime}(z,z^\prime)$ from the domain:
\begin{align}
{\cal{T}}^{-(2)}_{} = \Big\{ (z,z^\prime) \; ; \; z\in {\cal{T}}_{}^+, \; z^\prime \in {\cal{T}}_{}^- \Big\}\,, 
\end{align}
defined on ${M}_R^{(\mathbb{C})}\times {M}_R^{(\mathbb{C})}$. Notably, the permuted two-point function fulfills all the aforementioned requirements, as well.

After fulfilling the aforementioned requirements, the reconstruction theorem \cite{Streater} permits the construction of the corresponding QFT. Based on this, our current objective is to locate a doubled tensor-valued analytic function of the variable $(z, z^\prime)$ that demonstrates the mentioned properties.

\begin{widetext}
The explicit definition of the analytic two-point function $W_{\alpha\beta\alpha^\prime\beta^\prime}(z,z^\prime)$ can be derived from the provided solution (\ref{General plane wave}), using the following class of integral representations:
\begin{align}\label{two-point int.}
W_{\alpha\beta\alpha^\prime\beta^\prime}(z,z^\prime) = a^{2}_0 \int_\gamma \left(\frac{z\cdot\xi}{R}\right)^{-2} \left(\frac{z^\prime\cdot\xi}{R}\right)^{-2} \sum_{\kappa,\kappa^\prime=1}^3 \, {\cal E}^{\kappa\kappa^\prime}_{\alpha\beta}(z,\xi) \, {\cal E}^{\ast\kappa\kappa^\prime}_{\alpha^\prime\beta^\prime}(z^{\prime\ast},\xi)\; \mathrm{d}\mu^{}_\gamma(\xi)\,,
\end{align}
where the integration takes place over any orbital basis $\gamma$ of the future null-cone $C^+ \equiv \big\{ \xi\in \mathbb{R}^5 \; ; \; (\xi)^2 =0, \; \xi^0 > 0 \big\}$, $\mathrm{d}\mu^{}_\gamma$ denotes the intrinsic $C^+$ invariant measure on $\gamma$ and is derived from the Lebesgue measure of $\mathbb{R}^5$, and the normalization factor $a^{}_0$ is determined subsequently by applying the local Hadamard condition.

In order to determine whether the aforementioned conditions are satisfied by the analytic two-point function $W_{\alpha\beta\alpha^\prime\beta^\prime}(z,z^\prime)$, we should express the latter in a more explicit form. To accomplish this, we initiate with the analytic continuation of the polarization tensor (\ref{y 1}), and present it as follows:
\begin{align}\label{y 1'}
{\cal E}^{\kappa\kappa^\prime}_{\alpha\beta}(z,\xi) = \frac{1}{2} \Bigg( {\cal S} \,\varepsilon^\kappa_{\alpha}(z,\xi) \,\varepsilon^{\kappa^\prime}_{\beta}(z,\xi) - \frac{4}{9}\, \eta^{\kappa\kappa^\prime}_{} \left(\theta_{\alpha\beta} - \frac{D_{2\alpha} D_{1\beta}}{8R^2}\right) \Bigg)\,,
\end{align}
where we have used the fact that:
\begin{align}
\sum_{\kappa=1}^3\, \varepsilon^\kappa_{\alpha}(z,\xi) \, \varepsilon^{\kappa}_{\beta}(z,\xi) \left( \frac{z\cdot\xi}{R} \right)^{-2} = -\left(\theta_{\alpha\beta} - \frac{\overline{\xi}_\alpha \overline{\xi}_\beta}{(z\cdot\xi/R)^2}\right) \left( \frac{z\cdot\xi}{R} \right)^{-2} = - \frac{2}{3} \left(\theta_{\alpha\beta} - \frac{D_{2\alpha} D_{1\beta}}{8R^2}\right) \left( \frac{z\cdot\xi}{R} \right)^{-2}.
\end{align}
Then, by substituting the expression (\ref{y 1'}) into (\ref{two-point int.}), we simply develop the two-point function and obtain the following result:
\begin{align}\label{two-point int. 2}
W_{\alpha\beta\alpha^\prime\beta^\prime}(z,z^\prime) =& \frac{a^{2}_0}{4} \int_\gamma {\cal S}{\cal S}^\prime \Big(\sum_{\kappa=1}^3 \varepsilon^\kappa_{\alpha}(z,\xi) \,\varepsilon^{\ast\kappa}_{\alpha^\prime}(z^{\prime\ast},\xi) \Big) \Big(\sum_{\kappa^\prime=1}^{3} \varepsilon^{\kappa^\prime}_{\beta}(z,\xi) \, \varepsilon^{\ast\kappa^\prime}_{\beta^\prime}(z^{\prime\ast},\xi) \Big) \left(\frac{z\cdot\xi}{R}\right)^{-2} \left(\frac{z^\prime\cdot\xi}{R}\right)^{-2} \mathrm{d}\mu^{}_\gamma(\xi) \nonumber\\
& - \frac{16}{3} \left(\theta^{}_{\alpha\beta} - \frac{D^{}_{2\alpha} D^{}_{1\beta}}{8R^2}\right) \left(\theta^\prime_{\alpha^\prime\beta^\prime} - \frac{D^\prime_{2\alpha^\prime} D^\prime_{1\beta^\prime}}{8R^2}\right) c^{2}_0 \int_\gamma \left(\frac{z\cdot\xi}{R}\right)^{-2} \left(\frac{z^\prime\cdot\xi}{R}\right)^{-2} \mathrm{d}\mu^{}_\gamma(\xi) \,.
\end{align}
[Note that the primed operators only act on the primed coordinates, and similarly, the unprimed operators only act on the unprimed coordinates; then, for instance, we have $D_{2\alpha}D^{\prime}_{2\alpha^\prime}= D^{\prime}_{2\alpha^\prime}D_{2\alpha}$.] On the other hand, from the identity:
\begin{align}
&\sum_{\kappa=1}^3\, \varepsilon^\kappa_{\alpha}(z,\xi) \,\varepsilon^{\ast\kappa}_{\alpha^\prime}(z^{\prime\ast},\xi) = -\theta^{}_\alpha\cdot\theta^\prime_{\alpha^\prime} + \frac{(\theta^{}_\alpha\cdot z^\prime)\, \overline{\xi}_{\alpha^\prime}}{(z^\prime\cdot\xi)} + \frac{(\theta^{\prime}_{\alpha^\prime}\cdot z)\, \overline{\xi}_{\alpha}}{(z\cdot\xi)} + \frac{R^2 {\cal Z}\, \overline{\xi}_{\alpha} \overline{\xi}_{\alpha^\prime}}{(z\cdot\xi)(z^\prime\cdot\xi)}\,,
\end{align}
in which ${\cal Z} \equiv - z\cdot z^\prime/R^2$ is a dS-invariant length (see Appendix \ref{App. bitensors}), and also from the relation:
\begin{eqnarray}
R^{-2} D_{2\alpha} \,\varepsilon^{\kappa^\prime}_{\beta} \left( \frac{z\cdot\xi}{R} \right)^{-2} = \frac{-3\, {\cal S}\, \overline{\xi}_{\alpha}}{(z\cdot\xi)}\, \varepsilon^{\kappa^\prime}_{\beta} \left( \frac{z\cdot\xi}{R} \right)^{-2},
\end{eqnarray}
we can rewrite the analytic two-point function (\ref{two-point int. 2}) in the following form:
\begin{align}\label{two-point int. 3}
W(z,z^\prime) = \Delta(z,z^\prime) \widetilde{W}_1(z,z^\prime) + \Theta(z,z^\prime) \widetilde{W}_0(z,z^\prime)\,,
\end{align}
where the differential operators $\Delta(z,z^\prime)$ and $\Theta(z,z^\prime)$ respectively read as:
\begin{align}
\Delta(z,z^\prime) =& -\frac{9}{4}\, \bigg({\cal S} {\cal S}^\prime \,\theta\cdot\theta^\prime + \frac{{\cal S}\, R^{-2}(\theta\cdot z^\prime) D^\prime_2}{3} + \frac{{\cal S}^\prime\, R^{-2}(\theta^\prime\cdot z) D^{}_2}{3} - \frac{R^{-2} {\cal Z} D^{}_2 D^{\prime}_2}{9}\bigg), \\
\Theta(z,z^\prime) =& -\frac{16}{3} \left(\theta^{}_{} - \frac{D^{}_{2} D^{}_{1}}{8R^2}\right) \left(\theta^\prime_{} - \frac{D^\prime_{2} D^\prime_{1}}{8R^2}\right),
\end{align}
and the vector and scalar analytic two-point functions, denoted respectively by $\widetilde{W}_1(z,z^\prime)$ and $\widetilde{W}_0(z,z^\prime)$, are:
\begin{align}
\widetilde{W}_1(z,z^\prime) =& 4c^{}_0 \int_\gamma \sum_{\kappa=1}^3 \varepsilon^{\kappa}_{}(z,\xi) \, \varepsilon^{\ast\kappa}_{}(z^{\prime\ast},\xi) \left(\frac{z\cdot\xi}{R}\right)^{-2} \left(\frac{z^\prime\cdot\xi}{R}\right)^{-2} \mathrm{d}\mu^{}_\gamma(\xi) \nonumber\\
=& -4 \left( \theta\cdot\theta^\prime + \frac{R^{-2} (\theta\cdot z^\prime) D^\prime_1}{2} + \frac{R^{-2} (\theta^\prime\cdot z) D^{}_1}{2} - \frac{R^{-2} {\cal Z} D^{}_1D^\prime_1}{4} \right) \widetilde{W}_0(z,z^\prime)\,,\\
\label{scalar 2-point} \widetilde{W}_0(z,z^\prime) =& c^{}_0 \int_\gamma \left( \frac{z\cdot\xi}{R} \right)^{-2} \left( \frac{ z^\prime\cdot\xi }{R} \right)^{-2} \; \mathrm{d}\mu^{}_\gamma(\xi)\,.
\end{align}
Consequently, the analytic tensor two-point function can be expressed in terms of the scalar analytic two-point function as:
\begin{align}\label{two-point int. 4}
W_{\alpha\beta\alpha^\prime\beta^\prime}(z,z^\prime) = \mathfrak{D}_{\alpha\beta\alpha^\prime\beta^\prime}(z,z^\prime)\, \widetilde{W}_0(z,z^\prime)\,.
\end{align}
\end{widetext}

Note that the integral representation (\ref{scalar 2-point}) of the scalar analytic two-point function for well-chosen points $z,z^\prime\in{\cal{T}}^{+(2)}_{}$ in the domain determined by $( z-z^\prime )^2<0$, for instance, $z = \big( - \mathrm{i} R\cosh\varphi, - \mathrm{i} R\sinh\varphi,0,0,0 \big)$ and $z^\prime = \big( \mathrm{i} R,0,0,0,0 \big)$ (with $\varphi\geqslant 0$), yields \cite{Bros 2point func,Gazeau2022,GazeauPRL}:
\begin{align}\label{perikernel}
\widetilde{W}_0 \big( z,z^\prime \big) = \frac{-1}{8\pi^2 R^2} \frac{1}{1-{\cal Z}(z,z^\prime)}\,,
\end{align}
where the Hadamard condition has been used to establish a fixed value for the normalization factor $c^{}_0$.

Eventually, taking the boundary value of the analytic two-point function (\ref{two-point int. 4}) yields the corresponding Wightman two-point function:
\begin{align}
{\cal W}_{\alpha\beta\alpha^\prime\beta^\prime}(x,x^\prime) =&\, \text{bv}\; W_{\alpha\beta\alpha^\prime\beta^\prime}(z,z^\prime) \nonumber\\
=& \,\mathfrak{D}_{\alpha\beta\alpha^\prime\beta^\prime}(x,x^\prime)\; \Big(\text{bv}\; \widetilde{W}_0(z,z^\prime)\Big)\,,
\end{align}
with \cite{Bros 2point func,Gazeau2022,GazeauPRL}:
\begin{eqnarray}\label{2-point scalar}
\widetilde{{\cal W}}_0 ( x,x^\prime ) &=& \text{bv}\; \widetilde{W}_0 ( z,z^\prime ) \nonumber\\
&=& \frac{-1}{8\pi^2 R^2} \bigg(\mbox{\textbf{P}}\frac{1}{1-{\cal Z}(x,x^\prime)} \nonumber\\
&&- \mathrm{i}\pi\, \epsilon(x^0-x^{\prime 0}) \,\delta \big( 1-{\cal Z}(x,x^\prime) \big) \bigg)\,,\quad\quad
\end{eqnarray}
where $\mbox{\textbf{P}}$ stands for the principal part and $\epsilon(x^0-x^{\prime 0})=1,0,-1$ for $(x^0-x^{\prime 0}) >,=$, or $<0$, respectively. Note that the above expression corresponds precisely to the two-point function of the conformally coupled scalar field as presented in Ref. \cite{Tagirov}.

Here, it must be underlined that the derived kernel satisfies, through its construction, the previously mentioned conditions of indefinite sesquilinearity, \emph{covariance}\footnote{The covariance property arises from the group action on the dS modes (\ref{dS covariance K}) and the fact that the integral (\ref{two-point int.}) remains unaffected by the chosen orbital basis (for the latter property, see Ref. \cite{Bros 2point func, Gazeau2022}).}, \emph{locality}\footnote{To demonstrate the fulfillment of the locality requirement, one needs to utilize the identity $W^{}_{\alpha\beta\alpha^\prime\beta^\prime}(z,z^\prime) = W^{\ast}_{\alpha^\prime\beta^\prime\alpha\beta}(z^{\prime\ast},z^\ast)$, which can be easily checked through the integral representation (\ref{two-point int.}), and the relation $W^\ast_{\alpha^\prime\beta^\prime\alpha\beta}(z^{\prime\ast},z^\ast) = W^{}_{\alpha^\prime\beta^\prime\alpha\beta}(z^{\prime},z)$, which holds for spacelike separated points $z,z^\prime$ (obeying $( z-z^\prime )^2<0$). The latter relation is easy to be checked once we recall that the two-point function (\ref{two-point int. 4}), for $( z-z^\prime )^2<0$, takes the form:
$$W_{\alpha\beta\alpha^\prime\beta^\prime}(z,z^\prime) = \mathfrak{D}_{\alpha\beta\alpha^\prime\beta^\prime}(z,z^\prime)\,\left( \frac{-1}{8\pi^2 R^2} \frac{1}{1-{\cal Z}(z,z^\prime)}\right)\,, $$
where, by construction, $\mathfrak{D}(z,z^\prime) = \mathfrak{D}^\ast(z^\ast,z^{\ast\prime})$ and of course trivially ${\cal Z}(z,z^\prime) = {\cal Z}^\ast_{}(z^\ast,z^{\ast\prime})$. Therefore, for points $( z-z^\prime )^2<0$, we have:
$$W^{}_{\alpha\beta\alpha^\prime\beta^\prime}(z,z^\prime) = W^\ast_{\alpha^\prime\beta^\prime\alpha\beta}(z^{\prime\ast},z^\ast) = W^{}_{\alpha^\prime\beta^\prime\alpha\beta}(z^{\prime},z)\,.$$
Lastly, it should be observed that the spacelike separated pair $(x,x^\prime)$ resides within the same orbit of the complex dS group as the pairs $(z,z^\prime)$ and $(z^{\prime\ast},z^\ast)$. Consequently, the locality condition (\ref{locality re.}) is satisfied.}, index symmetrization, transversality, and \emph{normal analyticity}\footnote{The analytic characteristics of the tensor Wightman two-point function are derived from the representation of the dS tensor waves given in Eq. (\ref{masud}).}. These conditions are essential for obtaining a Wightman two-point function. It is also crucial to emphasize that the existence of a Wightman two-point function is a fundamental requirement in dS axiomatic QFT (see Ref. \cite{Gazeau2022}).

\subsection{Quantum field}
By having a comprehensive understanding of the Wightman two-point function ${\cal W}_{\alpha\beta\alpha^\prime\beta^\prime}(x,x^\prime)$, we can effectively employ the QFT formalism. It is expected that the tensor field $\hat{\cal K}_{\alpha\beta}(x)$ will serve as an operator-valued distribution on the spacetime manifold $M_R$, operating within the framework of a Hilbert space ${\cal H}$. In a more technical sense, we establish the definition of a vector-valued distribution, taking values in the space generated by the modes ${\cal K}_{\alpha\beta}(x,\xi) = \text{bv}\; {K}_{\alpha\beta}(z,\xi)$, for any test function $f_{\alpha\beta} \in D(M_R)$, as follows:
\begin{align}
x \;\mapsto\; p^{}_{\alpha\beta}(f)(x) =& \int_{M_R} {\cal W}_{\alpha\beta\alpha^\prime\beta^\prime}(x,x^\prime) f^{\alpha^\prime\beta^\prime}(x^\prime) \, \mathrm{d}\mu(x^\prime) \nonumber\\
=& \sum_{\kappa\kappa^\prime}\int_\gamma {\cal K}^{\kappa\kappa^\prime}_\xi(f) \,{\cal K}^{\kappa\kappa^\prime}_{\alpha\beta}(x,\xi)\, \mathrm{d}\mu_\gamma(\xi)\,,
\end{align}
where ${\cal K}^{\kappa\kappa^\prime}_\xi(f)$ refers to the smeared form of the modes:
\begin{align}
{\cal K}^{\kappa\kappa^\prime}_\xi(f) = \int_{M_R} {\cal K}^{\ast\kappa\kappa^\prime}_{\alpha\beta}(x,\xi) f^{\alpha\beta}(x) \, \mathrm{d}\mu(x)\,.
\end{align}
The space generated by the $p(f)$s is endowed with the indefinite invariant inner product:
\begin{align}
\langle p(f),p(g)\rangle =& \int_{M_R\times M_R} f^{\ast\alpha\beta}(x)\; {\cal W}_{\alpha\beta\alpha^\prime\beta^\prime}(x,x^\prime) \nonumber\\
&\hspace{1cm} \times\; g_{}^{\alpha^\prime\beta^\prime}(x^\prime) \, \mathrm{d}\mu(x) \mathrm{d}\mu(x^\prime)\,.
\end{align}

Then, the field, in the customary manner, is defined through the operator-valued distribution:
\begin{align}
\hat{\cal K}(f) = a\big(p(f)\big) + a^\dagger \big(p(f)\big)\,,
\end{align}
where the operators $a\big({\cal K}^{\kappa\kappa^\prime}(\xi)\big)$, denoted as $a^{\kappa\kappa^\prime}(\xi)$, and $a^\dagger\big({\cal K}^{\kappa\kappa^\prime}(\xi)\big)$, denoted as $a^{\dagger\kappa\kappa^\prime}(\xi)$, are respectively antilinear and linear in their respective arguments. On this basis, one obtains:
\begin{align}
\hat{\cal K}(f) =& \sum_{\kappa\kappa^\prime} \int_\gamma \Big({\cal K}^{\ast\kappa\kappa^\prime}_\xi(f) \, a_{}^{\kappa\kappa^\prime}(\xi) \nonumber\\
& \hspace{1.5cm} + {\cal K}^{\kappa\kappa^\prime}_\xi(f)\, a^{\dagger\kappa\kappa^\prime}_{}(\xi) \Big) \,\mathrm{d}\mu_\gamma(\xi)\,.
\end{align}
The unsmeared operator is:
\begin{align}\label{unsmeard field}
\hat{\cal K}_{\alpha\beta}(x) =& \sum_{\kappa\kappa^\prime} \int_\gamma \Big({\cal K}^{\kappa\kappa^\prime}_{\alpha\beta} (x,\xi) \, a_{}^{\kappa\kappa^\prime}(\xi) \nonumber\\
& \hspace{1.5cm} + {\cal K}^{\ast\kappa\kappa^\prime}_{\alpha\beta}(f)\, a^{\dagger\kappa\kappa^\prime}_{}(\xi) \Big) \,\mathrm{d}\mu_\gamma(\xi)\,.
\end{align}
where $a_{}^{\kappa\kappa^\prime}(\xi)$ fulfills the canonical commutation relations and is defined as follows:
\begin{align}
a_{}^{\kappa\kappa^\prime}(\xi)\, |0\rangle  = 0\,.
\end{align}
Note that the measure, denoted by $\mathrm{d}\mu_\gamma(\xi)$, exhibits the property of homogeneity as $\mathrm{d}\mu_\gamma(\ell\xi) = \ell^3 \mathrm{d}\mu_\gamma(\xi)$. Additionally, when considering ${\cal K}^{\kappa\kappa^\prime}_{\alpha\beta}(x,\ell\xi)$, it satisfies the homogeneity condition given by $\ell^{-2} {\cal K}^{\kappa\kappa^\prime}_{\alpha\beta}(x,\xi)$. Then:
\begin{align}
a_{}^{\kappa\kappa^\prime}(\ell\xi) \equiv a\big( {\cal K}^{\kappa\kappa^\prime}(\ell\xi) \big) = a\big( \ell^{-2}{\cal K}^{\kappa\kappa^\prime}(\xi) \big) = \ell^{-2} a_{}^{\kappa\kappa^\prime}(\xi)\,.
\end{align}
Moreover, one should notice that the integral representation (\ref{unsmeard field}) remains invariant regardless of the orbital basis $\gamma$, as elaborated in Refs. \cite{Bros 2point func, Gazeau2022}. Then, considering the hyperbolic type submanifold denoted as $\gamma_4$, which is defined as follows:\footnote{It is worth noting that the parametrization (\ref{fathi}) for $\xi$ exhibits an intriguing connection with the Poincar\'{e} massive UIRs. The null-vector $\xi \in C^+$ can actually be interpreted in terms of the four-momentum $(k^0,\vec{k})$ of a Minkowskian particle with mass $m$, as expressed below:
\begin{align}
\xi^\pm\, \big( \in \gamma^{}_4\big) = \Bigg( \frac{k^0}{m} = \sqrt{\frac{\vec{k}\cdot\vec{k}}{m^2} +1}, \frac{\vec{k}}{m}, \pm 1 \Bigg)\,, \nonumber
\end{align}
This parametrization allows us to satisfy the condition $(k^0)^2 - \vec{k}\cdot\vec{k} = m^2$, which corresponds to the mass-shell equation. This link between $\xi^\pm$ and the four-momentum enables a meaningful connection between the parametrization and the concept of particle mass \cite{Gazeau2022}.}
\begin{align}\label{fathi}
\gamma_4 = \Big\{ \xi\in {C}^+,\, \xi^4 = 1 \Big\} \bigcup \Big\{ \xi\in {C}^+,\, \xi^4 = -1 \Big\}\,,
\end{align}
the measure explicitly reads as $\mathrm{d}\mu_{\gamma_4}(\xi) = \mathrm{d}^3\vec{\xi}/\xi^0$, and consequently, the representation of the canonical commutation relations as:
\begin{align}
&\left[ a_{}^{\kappa\kappa^\prime}(\xi),\, a_{}^{\dagger\kappa^{\prime\prime}\kappa^{\prime\prime\prime}}(\xi^\prime) \right] \nonumber\\
& \hspace{1cm} = \left( \eta^{\kappa\kappa^{\prime\prime}}_{} \eta^{\kappa^\prime\kappa^{\prime\prime\prime}} + \eta^{\kappa\kappa^{\prime}}_{} \eta^{\kappa^{\prime\prime}\kappa^{\prime\prime\prime}} \right)\xi^0 \delta(\vec{\xi} - \vec{\xi}^\prime)\,.
\end{align}

The field commutation relations are:
\begin{align}
\left[ \hat{\cal K}_{\alpha\beta}(x),\, \hat{\cal K}_{\alpha^\prime\beta^\prime}(x^\prime) \right] =&\, 2\mathrm{i}\, \text{Im}\, \langle p_{\alpha\beta}(x),\, p_{\alpha^\prime\beta^\prime}(x^\prime)\rangle \nonumber\\
=&\, 2\mathrm{i}\, \text{Im}\, {\cal W}_{\alpha\beta\alpha^\prime\beta^\prime}(x,x^\prime) \nonumber\\
=&\, 2\mathrm{i}\, \mathfrak{D}_{\alpha\beta\alpha^\prime\beta^\prime}\, \text{Im}\, \widetilde{\cal W}_0(x,x^\prime)\,,
\end{align}
where we have used Eq. (\ref{two-point int. 4}) while, following the relation (\ref{2-point scalar}), we have:
\begin{align}
\text{Im}\, \widetilde{\cal W}_0(x,x^\prime) = \frac{1}{8\pi R^2} \,\epsilon(x^0-x^{\prime 0}) \, \delta \big( 1-{\cal Z}(x,x^\prime)\big)\,.
\end{align}
Finally, we have the commutator:
\begin{align}\label{light-cone propagation}
&\mathrm{i}\, G_{\alpha\beta\alpha^\prime\beta^\prime} (x,x^\prime) = \left[ \hat{\cal K}_{\alpha\beta}(x),\, \hat{\cal K}_{\alpha^\prime\beta^\prime}(x^\prime) \right] \nonumber\\
&\hspace{0.4cm} = \frac{\mathrm{i}}{4\pi R^2}\,  \mathfrak{D}_{\alpha\beta\alpha^\prime\beta^\prime}\, \epsilon(x^0-x^{\prime 0}) \, \delta \big( 1-{\cal Z}(x,x^\prime)\big)\,.
\end{align}

\emph{\textbf{Important remark:}} It is worth noting that the right-hand side of the above equation clearly demonstrates the light-cone propagation of the dS partially massless graviton field.

\section{Conclusion and outlook} \label{Sec. Conclusion}
Over the past four decades, there has been a remarkable proliferation of diverse and sometimes conflicting approaches to dS physics. From the vantage point of the mathematical physics community, dS spacetime holds a privileged status as the unique maximally symmetric solution to the Einstein equation with a positive cosmological constant. It encompasses a group of motions characterized by $10$ essential parameters and facilitates the identification of conventional observables like mass and spin. This unique process of identification owes its feasibility to the essence of the dS group, which smoothly transforms into the Poincar\'{e} group through a procedure known as the contraction procedure. As a result, dS spacetime stands as an invaluable laboratory for constructing a coherent formulation of elementary systems in the sense defined by Wigner, intricately connected with the UIRs of the relativity group. In the realm of this captivating framework, one is fundamentally engaged with the entirety of dS spacetime's global structure. [For a more comprehensive understanding of this approach, see Ref. \cite{Gazeau2022} and the cited references therein.]

From the perspective of cosmological physics, dS spacetime is not merely a theoretical solution but a dynamic model capable of describing the Universe's late-time evolution under the influence of dark energy. Equally significant is its role in portraying the inflationary epoch during the early moments of the Universe. In this context, in contrast to the former approach to dS physics, the concept of symmetry breaking assumes a pivotal role within the framework of dS physics, strictly speaking, the inflationary scenario. Within the inflationary model, symmetry breaking is frequently associated with the behavior of a scalar field known as the ``inflaton''. Initially existing in a symmetric state, this field undergoes a phase transition as the Universe expands and cools during inflation, resulting in the breaking of its symmetry and giving rise to the particles and forces we observe in the Universe. In this landscape, researchers often operate within a coordinate system - frequently the dS flat spacetime coordinates - that may not encompass the entirety of dS spacetime's global structure. [We refer readers for more details to Refs. \cite{Goodhew, mm, nn, ppp, aa, bb, cc, dd, ee, ff}, for instance.]

The present study should be recognized as an endeavor within the former perspective. Its primary objective is to present a coherent formulation of elementary systems - specifically the partially massless graviton field - within dS spacetime.

Pursuing this aim, we have successfully quantized the partially massless graviton field in dS spacetime,  associated with the discrete series UIR $\Pi^\pm_{2,1}$, by applying various established techniques from previous studies. These include utilizing the ambient space formalism, constructing appropriate modes, ensuring dS covariance and Gupta-Bleuler triplets, constructing the Wightman two-point function, and ultimately achieving covariant quantization of the field. One immediate consequence of this construction validates the widely accepted notion regarding the light-cone propagation behavior of the partially massless graviton field in dS spacetime.

Our calculations have uncovered the presence of two distinct types of logarithmic singularities inherent to the theory. It is imperative to emphasize that these singularities manifest exclusively within the nonphysical domain of the theory, thus carrying no impact on the underlying physics of the problem.

Given this context, a logical sequence urges us to investigate the theory's stability. This pursuit can be realized through the lens of first-order perturbation theory within the framework of an interacting QFT. To be precise, this progression involves a meticulous examination of whether the partially massless graviton, as described by the theory, undergoes decay in a dS spacetime. Technically speaking, this endeavor mandates a comprehensive exploration of the decomposition of the tensor product associated with the relevant UIRs, as outlined in Ref. \cite{Lifetime} (see section \ref{Subsec. Lifetime}). This task indeed entails intricate mathematical calculations at its core, constituting a substantial avenue for future exploration.

\setcounter{equation}{0} \section*{Acknowledgements}
Hamed Pejhan is supported by the Bulgarian Ministry of Education and Science, Scientific Programme ``Enhancing the Research Capacity in Mathematical Sciences (PIKOM)", No. DO1-67/05.05.2022. Furthermore, Hamed Pejhan would like to express sincere appreciation to Kiril Hristov for his invaluable comments.

\section{Appendices}

\begin{appendix}

\setcounter{equation}{0} \section{Plane-wave type solutions} \label{App. Plane Waves}
As mentioned earlier, the QFT construction outlined in this paper relies technically on dS plane waves, which have a global definition on the dS hyperboloid $M_R$. In this appendix, we provide a concise overview of this concept, focusing on dS scalar waves to avoid unnecessary technical intricacies while preserving the fundamental concepts.

When examining the scalar wave equation(s) (\ref{Wave Eq. scalar}) within the specific allowable ranges for the unified complex parameter $\tau$, a continuous set of simple solutions, commonly known as dS plane waves, becomes prominent. These waves explicitly reads as \cite{Bros 2point func,GazeauPRL}:
\begin{align}\label{CC solution}
\phi(x) = \left(\frac{x\cdot\xi}{R}\right)^\tau ,
\end{align}
where $x$ and $\xi$ respectively live in $M_R$ and in the null-cone $C$ in $\mathbb{R}^5$:
\begin{align}
C = \Big\{ \xi\in \mathbb{R}^5 \; ; \; (\xi)^2 \equiv \xi\cdot\xi = \eta^{}_{\alpha\beta}\xi^\alpha\xi^\beta =0 \Big\}\,.
\end{align}

Here, without delving into mathematical intricacies, we outline a couple of crucial points regarding the dS waves.

\emph{First:} The dS plane waves (\ref{CC solution}) on the null-cone $C$ can be completely determined by their values on a well-chosen curve, referred to as the orbital basis $\gamma\subset C$. This is because these waves, as functions of $\xi$, are homogeneous with the degree of homogeneity $\tau$. [It is worth noting that the dS plane waves are also homogeneous with the homogeneity degree $\tau$ on the dS hyperboloid $M_R \subset\mathbb{R}^5$, which is characterized as a (pseudo-)sphere with $x\cdot x = -R^2$ (where $R$ is a constant). But, as functions of $\mathbb{R}^5$, while they remain homogeneous, their degree of homogeneity becomes zero; since, in this case, $R$ must be treated as a function of $x$, i.e., $R(x)= - \sqrt{- x\cdot x}$.]

\emph{Second:} The dS waves, when viewed as \emph{functions} on ${M}_R$, exhibit multivaluedness and are only locally defined on connected open subsets of ${M}_R$. The reason for their multivaluedness is that the expression $x\cdot\xi$ can give rise to negative values. They have local definitions because they exhibit singularity on certain lower-dimensional subsets of ${M}_R$, such as the spatial boundaries $x^0 = \pm x^4 \Leftrightarrow (x^1)^2 + (x^2)^2 + (x^3)^2 = R^{2}$. To elaborate, recalling $\mbox{Re}(\tau)<0$ (see subsequent discussion to Eq. (\ref{Wave Eq. scalar})), one may associate the boundaries with $\xi=(\xi^0=\pm\xi^4,0,0,0,\xi^4) \in C$, where $x\cdot\xi = 0$.

\emph{Third:} By treating the dS waves as \emph{distributions}, however, a single-valued global definition can be attained \cite{Bros 2point func,GazeauPRL}. This involves considering them as the boundary values \emph{(in the distribution sense)} of the analytic continuations of solutions to the respective field equation, to appropriate domains within the complexified dS manifold:
\begin{align}\label{complex dS}
{M}_R^{(\mathbb{C})} &\equiv& \Big\{ z=x+ \mathrm{i} y \in {\mathbb{C}}^5 \;;\; (z)^2 = \eta^{}_{\alpha\beta} z^\alpha_{} z^\beta_{} = -R^2 \Big\} \,,
\end{align}
where ${\mathbb{C}}^5$ refers to the ambient complex Minkowski spacetime. The minimal domains of analyticity required to obtain a single-valued global definition of the waves are found to be the ``forward" and ``backward tubes" of ${M}_R^{(\mathbb{C})}$, respectively defined by \cite{Bros 2point func, GazeauPRL}:
\begin{eqnarray}\label{forward and back tube dS}
{\cal{T}}^\pm = \Big\{ \mathbb{R}^5 + \mathrm{i} \interior{V}^\pm \Big\} \cap {M}_R^{(\mathbb{C})}\,,
\end{eqnarray}
where, as mentioned earlier, the domains ${\interior{V}}^\pm \equiv \big\{y \in\mathbb{R}^5 \;;\; (y)^2 > 0, y^0 \gtrless 0 \big\}$ originate from the causal structure of ${M}_R$. As such, as far as $\xi$ is restricted to the future null-cone $C^+ \equiv \big\{ \xi\in \mathbb{R}^5 \; ; \; (\xi)^2 =0, \; \xi^0 > 0 \big\}$, the imaginary part of the resulting complexified waves $(z\cdot\xi/R)^\tau$ consistently maintains a specific sign, and moreover, $z\cdot\xi \neq 0$. The former property ensures the single-valued determination of $(z\cdot\xi/R)^\tau$ as follows:
\begin{align}
\left(\frac{z\cdot\xi}{R}\right)^\tau = \exp \left(\tau \left[\mathrm{i} \; \mbox{arg}\left(\frac{z\cdot\xi}{R}\right) + \log\left|\frac{z\cdot\xi}{R}\right| \right] \right)\,,
\end{align}
where $\mbox{arg} \big({z\cdot\xi}/{R}\big) \in \; ]-\pi,\pi[$.

\begin{widetext}
Then, taking the boundary value \emph{(in the distribution sense)} of the above-complexified waves, we obtain the single-valued global plane wave reading of the general solution (\ref{CC solution}):
\begin{align}
\left(\frac{x\cdot\xi}{R}\right)^\tau = \text{bv}\; \left(\frac{z\cdot\xi}{R}\right)^\tau = \left(\frac{(x+\mathrm{i}y)\cdot\xi}{R}\right)^{\tau}\bigg|_{\xi \in C^+,\; y \in \interior{{V}}^+, \; y \rightarrow 0}\,,\quad
\end{align}
where the last term, which symbolically represents the boundary values \emph{(in the distribution sense)} of analytic continuation to the forward-tube ${\cal{T}}^+$ of the scalar waves (\ref{CC solution}), explicitly reads:
\begin{eqnarray}\label{Fourier x.xi}
\phi^{+}_{\tau,\xi}(f) &=& \int_{{M}_R} c \; \left( \frac{(x+ \mathrm{i} y)\cdot\xi}{R} \right)^\tau \bigg|_{\xi \in C^+,\; y \in \interior{{V}}^+, \; y \rightarrow 0} \; f(x) \; \mathrm{d}\mu(x) \nonumber\\
&=& \int_{{M}_R} \underbrace{\Bigg( c \; \left[ \vartheta\left(\frac{x\cdot\xi}{R}\right) + \vartheta\left(-\frac{x\cdot\xi}{R}\right) e^{+ \mathrm{i} \pi\tau} \right] \; \left|\frac{x\cdot\xi}{R}\right|^\tau \Bigg)}_{\equiv \phi^{+}_{\tau,\xi}(x)} \; f(x) \; \mathrm{d}\mu(x)\,, \quad \tau=-1,-2\,.
\end{eqnarray}
where, in this ``Fourier transform", $f(x)$ belongs to $\mathfrak{D}({M}_R)$, which is the space of infinitely differentiable functions with compact support on ${M}_R$, $\mathrm{d}\mu(x)$ refers to the invariant measure on ${M}_R$, $\vartheta$ to the Heaviside function, and $c$ to a real-valued constant that is determined by applying the local Hadamard condition to the corresponding two-point function.
\end{widetext}

\emph{Lastly, the fourth and final point to consider:} The dS plane waves are not square integrable with respect to the Klein-Gordon inner product or else (!?). However, they serve as generating functions for physically meaningful dS entities, such as square-integrable eigenfunctions of the dS quadratic Casimir operator. [Recall that these square-integrable eigenfunctions give rise to (projective) Hilbert spaces carrying the dS UIRs.] In this sense, within dS relativity, the aforementioned plane waves exhibit similarities to the conventional waves in Minkowskian or Galilean quantum mechanics, which, by superimposing nonsquare-integrable plane waves, one can construct physical wave functions (wave packets) in a similar manner.

For a comprehensive review of the above content, readers are referred to Ref. \cite{Gazeau2022}.

\setcounter{equation}{0} \section{Two-point function from maximally symmetric bitensors in ambient space} \label{App. bitensors}
In this appendix, we establish a connection between our construction and the \emph{maximally symmetric bitensors}\footnote{By their very definition, bitensors, which are functions of two points $x$ and $x^\prime$, exhibit tensor-like transformation properties at each point. When these bitensors maintain the invariance of dS spacetime, they are known as maximally symmetric.} that were introduced by Allen and Jacobson in Ref. \cite{AllenJacobson}.

Let $\sigma(x,x^\prime)$ denote the (pseudo-)distance on $M_R$, that is, the distance along the geodesic that links the two points/events $x,x^\prime \in M_R$. [It is important to note that if there is no geodesic connecting $x$ and $x^\prime$, the geodesic distance is determined by a unique analytic extension.] Technically, $\sigma(x,x^\prime)$ is implicitly given for timelike separated points $x,x^\prime \in {M}_R$ by \cite{Bros 2point func}:
\begin{eqnarray}
\cosh \left(\frac{\sigma(x,x^\prime)}{R}\right) = - \frac{x\cdot x^\prime}{R^2} \equiv {\cal{Z}}\,,
\end{eqnarray}
and for spacelike separated points $x,x^\prime \in {M}_R$, such that $| x\cdot x^\prime | < R^2$, by \cite{Bros 2point func}:
\begin{eqnarray}
\cos \left(\frac{\sigma(x,x^\prime)}{R}\right) = - \frac{x\cdot x^\prime}{R^2} \equiv {\cal{Z}}\,.
\end{eqnarray}

Any maximally symmetric bitensor living in dS spacetime, with respect to the lines sketched in Ref. \cite{AllenJacobson} by Allen and Jacobson, can be expressed as a sum of products of the following three basic tensors:
\begin{eqnarray}
&n_\mu = \nabla_\mu \sigma(x, x^{\prime})\,,\quad 
n_{\mu^{\prime}} = \nabla_{\mu^{\prime}} \sigma(x,x^{\prime})\,,&\nonumber\\
&g_{\mu\nu^{\prime}} = -c^{-1}({\cal{Z}})\nabla_{\mu}n_{\nu^{\prime}}+n_\mu n_{\nu^{\prime}}\,,&
\end{eqnarray}
where $c({\cal{Z}})= {1}/{R\sqrt{1-{\cal{Z}}^2}}$, such that the expansion coefficients are characterized in terms of the geodesic distance $\sigma(x,x^\prime)$. In this context, a dS rank-$2$ two-point function takes the form:
\begin{eqnarray}
{\cal{W}}_{\mu\nu\mu^{\prime}\nu^{\prime}} &=& {\cal{A}}_1 (\sigma)g_{\mu\nu}g^{\prime}_{\mu^{\prime}\nu^{\prime}} + {\cal{A}}_2 (\sigma)g_{\mu\mu^{\prime}}g^{\prime}_{\nu\nu^{\prime}}\nn\\
&&+ {\cal{A}}_3 (\sigma)(g_{\mu\nu}n_{\mu^{\prime}}n_{\nu^{\prime}} + g^{\prime}_{\mu^{\prime}\nu^{\prime}}n_\mu n_\nu)\nn\\
&&+ {\cal{A}}_4 (\sigma)g_{\mu\mu^{\prime}}n_\nu n_{\nu^{\prime}} + {\cal{A}}_5 (\sigma)n_\mu n_\nu n_{\mu^{\prime}}n_{\nu^{\prime}}\,.\quad\quad
\end{eqnarray}

In terms of the ambient space notations employed in this paper, the three basic bitensors $n_\mu$, $n_{\mu^\prime}$, and $g_{\mu\nu^\prime}$ respectively correspond to:
\begin{eqnarray}
\overline{\partial}_\alpha \sigma(x,x^{\prime})\,,\;\;\;{\overline{\partial}}_{\beta^{\prime}}^{\prime} \sigma(x,x^{\prime})\,,\;\;\;\theta_\alpha\cdot\theta^{\prime}_{\beta^{\prime}}\,.
\end{eqnarray}
To verify this, one just needs to take into account the restriction to the hyperboloid described by Eq. (\ref{tt}):
\begin{itemize}
\item{When $ {\cal{Z}}=\cos(\sigma/R)$:
\begin{eqnarray}
n_\mu &=& x^{\alpha}_{\,\,,\,\mu}\; \overline{\partial}_\alpha \sigma(x,x^{\prime})= c({\cal{Z}})\; x^{\alpha}_{\,\,,\,\mu} (x^{\prime}\cdot\theta_\alpha)\,,\\
n_{\nu^{\prime}} &=& {x^{\prime}}^{\beta^{\prime}}_{\,\,,\,\nu^{\prime}}\; \overline{\partial}_{{\beta^{\prime}}}^{\prime} \sigma(x,x^{\prime}) = c({\cal{Z}})\; {x^{\prime}}^{\beta^{\prime}}_{\,\,,\,\nu^{\prime}} (x\cdot\theta^{\prime}_{\beta^{\prime}})\,,\quad\quad\\
\nabla_\mu n_{\nu^{\prime}} &=& x^{\alpha}_{\,\,,\,\mu}{x^{\prime}}^{\beta^{\prime}}_{\,\,,\,\nu^{\prime}}\; \theta^\varrho_\alpha{\theta^{\prime}}^{\gamma^{\prime}}_{\beta^{\prime}}\; \overline{\partial}_\varrho\overline{\partial}_{{\gamma^{\prime}}}^{\prime} \sigma(x, x^{\prime}) \nonumber\\
&=& c({\cal{Z}})\Big[x^{\alpha}_{\,\,,\,\mu}{x^{\prime}}^{\beta^{\prime}}_{\,\,,\,\nu^{\prime}}\; (\theta_\alpha \cdot\theta^{\prime}_{\beta^{\prime}}) - n_\mu n_{\nu^{\prime}}{\cal{Z}}\Big]\,,
\end{eqnarray}
where we recall from section \ref{Subsec. dS field equ.} that ${x}^{\alpha}_{\,\,,\,\mu} = {\partial{x}^{\alpha}}/{\partial {X^{\mu}}}$ and ${x^{\prime}}^{\beta^{\prime}}_{\,\,,\,\nu^{\prime}} = {\partial {x^{\prime}}^{\beta^{\prime}}}/{\partial {X^{\prime}}^{\nu^{\prime}}}$.}

\item{When ${\cal{Z}}=\cosh (\sigma/R)$, $c({\cal{Z}})$, $n_\mu$, and $n_{\nu^{\prime}}$ are multiplied by $\mathrm{i}$. As such, for both cases, we have:
\begin{eqnarray}
g_{\mu\nu^{\prime}}+({\cal{Z}}-1)n_\mu n_{\nu^{\prime}}=x^{\alpha}_{\,\,,\,\mu}{x^{\prime}}^{\beta^{\prime}}_{\,\,,\,\nu^{\prime}} (\theta_\alpha \cdot\theta^{\prime}_{\beta^{\prime}})\,.
\end{eqnarray}}
\end{itemize}

\end{appendix}

\end{document}